\documentstyle[11pt]{article}
\def\div{{\,\rm div \,}}
\def\g{{\,\rm \gamma \,}}
\def\sign{{\,\rm sign\, }}

\def\T{{\cal T}}

\def\CC{{\,\rm C\,}}

\def\sym{{\,\rm sym \,}}

\def\ii{{\,\rm i \,}}

\def\B{{\,\cal B \,}}

\def\+M{{\,\rm M^{n\times n}_+ \,}}
\def\tr{{\,\rm tr \,}}

\def\qfq{{\quad\mbox{for}\quad}}

\def\ii{{\,\rm i \,}}

\def\lam{\lambda}

\def\P{{\cal P}}

\def\N{{\cal N}}

\newfont{\Blackboard}{msbm10 scaled 1200}

\newfont{\roma}{cmr10 scaled 1200}

\def\<{{\langle}}
\def\>{{\rangle}}

\def\g{\gamma}
\def\Ga{\Gamma}

\def\si{\sigma}

\def\a{\alpha}
\def\b{\beta}

\def\Om{\Omega}

\newtheorem{thm}{{}\hskip\parindent Theorem}[section]
\newtheorem{lem}{{}\hskip\parindent Lemma}[section]
\newtheorem{pro}{{}\hskip\parindent Proposition}[section]

\newtheorem{cor}{{}\hskip\parindent Corollary}[section]

\newtheorem{rem}{{}\hskip\parindent Remark}[section]

\def\dfrac{\displaystyle\frac}

\def\pl{\partial}
\def\rw{\rightarrow}

\def\na{\nabla}

\def\be{\begin{equation}}
\def\ee{\end{equation}}
\def\beq{\arraycolsep=1.5pt\begin{eqnarray}}
\def\eeq{\end{eqnarray}}

\def\P{\cal P}
\def\R{I\!\!R}

\def\n{\vec{n}}
\large
\title{Linear Strain Tensors and Optimal Exponential of thickness in Korn's Inequalities for Hyperbolic  Shells}
\date{}
\author{
Peng-Fei YAO\\[0.3cm]
Key Laboratory of  Systems and Control\\
Institute of Systems Science,
Academy of Mathematics and Systems Science\\
Chinese Academy of Sciences, Beijing 100190, P. R. China\\
School of Mathematical Sciences\\
University of Chinese Academy of Sciences, Beijing 100049, China\\
e-mail: pfyao@iss.ac.cn}

\begin{document}
\maketitle
 \footnote{This work is  supported by the National
Science Foundation of China, grants  no. 61473126 and no. 61573342, and Key Research Program of Frontier Sciences, CAS, no. QYZDJ-SSW-SYS011.}

\begin{quote}
\begin{small}
{\bf Abstract} \,\,\,We perform a detailed analysis of the solvability of linear strain equations
on hyperbolic surfaces to obtain $L^2$ regularity solutions. Then the rigidity results on the strain tensor of the middle surface are implied by the $L^2$ regularity for non-characteristic regions. Finally, we obtain the optimal
constant in the first Korn inequality scales like $h^{4/3}$ for hyperbolic shells, generalizing the  assumption that the middle surface of the shell is given by a single
principal system in the literature.
\\[3mm]
\end{small}
\end{quote}

\setcounter{equation}{0}
\section{Introduction}
\def\theequation{1.\arabic{equation}}
\hskip\parindent The goal of the present paper is twofold to study the solvability of linear strain equations and the optimal constant in the first Korn inequality  for hyperbolic shells.

The Linear strain equations plays a fundamental role in the theory of thin shells, see \cite{HoLePa, LePa,LeMoPa, LeMoPa1, Yao2017}. The solvability of the strain equation is needed to prove the density of smooth infinitesimal isometries  in the $W^{2,2}(\Om,\R^3)$ infinitesimal isometries and and the matching property of the smooth enough infinitesimal isometries  with higher
order infinitesimal isometries \cite{HoLePa, LeMoPa1, Yao2017}. This ¡±matching property¡± is an important tool in
obtaining recovery sequences ($\Ga$-lim sup inequlity) for dimensionally-reduced shell theories
in elasticity, when the elastic energy density scales like $h^\b,$ $\b\in(2, 4),$ that is, intermediate
regime between ¡±pure bending¡± ($\b=2$) and the von-K¡äarm¡äan regime ($\b=4$). Such results have been obtained for elliptic surfaces \cite{LeMoPa1}, developable surfaces \cite{HoLePa}, and hyperbolic surfaces \cite{Yao2017}, respectively.
A survey on this topic is presented in \cite{LePa}.

Here we present a direct method of solving the linear strain equations for the hyperbolic middle surface, different from \cite{Yao2017}. The present approach  is relatively simple and allows us to obtain a lower regularity on the solution, see Theorem \ref{t1} later. Fortunately, this regularity implies the  rigidity results on the strain tensor of the middle surface which is one of the key ingredients for
 the optimal constant in the first Korn inequality  for hyperbolic shells (\cite{Ha2}).

Originally, Korn's inequalities were used to prove existence, uniqueness and well-posedness of
boundary value problems of linear elasticity (see e.g., \cite{Ci,Lo}). The optimal exponential of thickness in Korn's inequalities for thin shells represents the relationship between the rigidity and the thickness of a shell when the small deformations take place since  Korn's inequalities are linearized from the geometric rigidity inequalities under the small deformations (\cite{FrJaMu}). Thus it is the best Korn constant in the Korn inequality that is of
central importance (e.g., \cite{CiOlTr,LeMu, Na, Na1, PaTo, PaTo1}). Moreover, it is  ingenious that the best Korn constant is subject to the Gaussian curvature. The one for the parabolic shell  scales like $h^{3/2}$ (\cite{GH,GH1}), for the hyperbolic shell, $h^{4/3}$ (\cite{Ha2}) and for the elliptic shell, $h$ (\cite{Ha2}). All those results were derived under the main assumption that the middle surface of the shell is given by a single principal coordinate system in order to carry out some necessary computation. This assumption is
\be S=\{\,{\bf r}(z,\theta)\,|\,(z,\theta)\in[1,1+l]\times[0,\theta_0]\,\},\label{as}\ee where the properties
$$\nabla_{\pl z}\n=\kappa_z\pl z,\quad\nabla_{\pl\theta}\n=\kappa_\theta\pl\theta\qfq p\in S $$ hold.

In the case of the parabolic or hyperbolic shell, a principal coordinate only exists locally (\cite{Yao2018}). There is even no such a local existence for the elliptic shell. However, the  assumption (\ref{as}) in \cite{GH,GH1, Ha2} can be removed if the Bochner technique is employed to perform  some necessary computation. The Bochner technique
provides us the great simplification in computation, for example, see \cite{Ho} or \cite{Yao2011}.  It has been done in the cases of the parabolic and elliptic shells in \cite{Yao2018}. Here we treat the
hyperbolic shell by combining the rigidity lemma of the strain tensor of the middle surface, given in this paper,  and the interpolation inequality \cite{Ha3} to obtain that the best constant in Korn's
inequality scales like $h^{4/3},$ removing the assumption (\ref{as}).

Let $M\subset\R^3$ be a $\CC^3$ surface with the induce metric $g$ and a normal field $\n.$ Let $S\subset M$ be an open bounded set with a regular boundary $\pl S.$  Suppose that $S$ is the middle surface of the shell with thickness $h>0$
$$\Om=\{\,x+t\n(x)\,|\,x\in S,\,-h/2<t<h/2\,\}.$$
A shell $\Om$ is said to be hyperbolic if
$$\kappa(p)<0\qfq p\in\overline{S},$$ where $\kappa$ is the Gaussian curvature. Throughout the paper $\Om$ is assumed to be hyperbolic.

Let $y\in H^1(S,\R^3)$ be a displacement of the middle surface $S.$ We decompose $y$ as
$$y=W+w\n,\quad w=\<y,\n\>,$$ where $\<\cdot,\cdot\>$ denotes the dot metric of the Euclidean space $\R^3.$ The (linear) strain tensor of the middle surface (related to the displacement $y$) is defined by $$\Upsilon(y)=\sym DW+w\Pi,$$ where $D$ is the Levi-Civita connection of the induced metric $g$ on $S,$
$$\sym DW=\frac12(DW+DW^T),$$ and
$$\Pi(\a,\b)=\<\nabla_\a\n,\b\>\qfq\a,\,\,\b \in M_x,\quad x\in S$$ is the second fundamental form of surface $M.$ $y\in H^1(S,\R^3)$ is said to be an infinitesimal isometry if
$$\Upsilon(y)=0.$$
For $U\in L^2(S,T^2)$ given, consider problem
\be\Upsilon(y)=U\qfq p\in S.\label{s}\ee

We say that $S$ is a {\it non-characteristic region} if one of the following assumptions (I)--(IV) holds.

(I)\,\,\,Let
\be S=\{\,\a(t,s)\,|\,(t,s)\in(0,a)\times(0,b)\,\},\label{main1}\ee where $\a:$ $[0,a]\times[0,b]\rw M$ is an imbedding map which is a family of regular curves with two parameters $t,$ $s$ such that
\be \Pi(\a_t(t,s),\a_t(t,s))\not=0,\quad\mbox{for all}\quad (t,s)\in[0,a]\times[0,b],\label{main2}\ee
\be \Pi(\a_s(0,s),\a_s(0,s))\not=0,\quad \Pi(\a_s(a,s),\a_s(a,s))\not=0,\quad\mbox{for all}\quad s\in[0,b],\label{main3}\ee
\be \Pi(\a_t(0,s),\a_s(0,s))=\Pi(\a_t(a,s),\a_s(a,s))=0,\quad\mbox{for all}\quad s\in[0,b].\label{main4}\ee

(II)\,\,\,Let $\a(\cdot,s)$ be a closed curve with the period $a$ for each $s\in[0,b]$ given. Let
\be S=\{\,\a(t,s)\,|\,t\in[0,a),\,\,s\in[0,b]\,\},\label{main5}\ee where $\a:$ $[0,a)\times[0,b]\rw M$ is an imbedding map if $\a(\cdot,b)$ is a closed curve;
$\a:$ $[0,a)\times[0,b)\rw M$ is an imbedding map if $\a(\cdot,b)$ is one point. Moreover,
for each $s\in[0,b],$
$$\Pi(\a_t(t,s),\a_t(t,s))\not=0\qfq t\in[0,a].$$

(III)\,\,\,Let $S$ be given in (\ref{main1}) such that (\ref{main3}) and (\ref{main4}) hold. Let $m\geq2$ be an integer. We assume that for each $s\in[0,b]$ given, the curve $\a(\cdot,s)$ consists of
$m$ $\CC^1$ non-characteristic curves, i.e., there are $t_0=0<t_1<\cdots <t_{m-1}<t_m=a,$ such that $\a(t,s)$ are $\CC^1$ non-characteristic curves for $t\in[t_i,t_{i+1}],$ respectively, for $i=0,$ $1,$ $\cdots,$  and $m-1.$ Moreover, the curves $\zeta_i(s)=\a(t_i,s):$ $[0,b]\rw M$ are supposed to be extended to $[-\varepsilon,b]$ satisfying
\be\zeta_i(s)\not\in S\qfq s\in[-\varepsilon,0],\quad 1\leq i\leq m-1,\label{H1.8}\ee such that  $\zeta_i$ are  non-characteristic curves in $s\in[-\varepsilon,b]$ where $\varepsilon>0$ is given small.
In addition, at all the connection points $\a(t_i,s)$ one of the connection conditions ({\bf H1})--({\bf H4}) below holds where $\b(t)=\a(t_i+t-\varepsilon,s)$($\varepsilon>0$ small), $\g(t)=\a(t_k+t,s),$ and $\zeta(t)=\a(t_k,s+t)$ for $1\leq i\leq m-1$ and $s\in[0,b].$

(IV)\,\,\, Let $S$ be given in (\ref{main5}) and $m\geq2$ be an integer. For each $s\in[0,b],$ we assume that the closed curve $\a(t,s)$ consists of $m$ $\CC^1$ non-characteristic curves, i.e.,
there are $t_1=0<\cdots <t_{m-1}<t_{m}=a,$ such that $\a(t,s)$ are $\CC^1$ non-characteristic curves for $t\in[t_i,t_{i+1}],$ respectively, for $i=1,$ $2,$ $\cdots,$ and $m-1.$ Moreover, assumptions (\ref{H1.8}) hold. In addition, at all the connection points one of the connection conditions ({\bf H1})--({\bf H4}) below holds as in (III).

In the case of $(III),$ or $(IV),$ to solve (\ref{s}) under certain corresponding boundary data, we further need some connection conditions at the connection points.

{\bf Connection conditions}\,\,\,Let $\b:$ $[0,\varepsilon]\rw M,$  $\g:$ $[0,\varepsilon]\rw M,$ and $\zeta:$ $[-\varepsilon,\varepsilon]\rw M$ be  noncharacteristic  curves with $\b(\varepsilon)=\g(0)=\zeta(0)$ where $\varepsilon>0$ is small. We say that $\b,$ $\g,$ and $\zeta$ satisfy the {\it connection condition} $({\bf H}k)$ at $p=\b(\varepsilon)$ if the following assumption $({\bf H}k)$ holds true where $1\leq k\leq4.$
$${\bf (H1)}\quad\quad
\Pi(\b'(\varepsilon),\b'(\varepsilon))\Pi(\g'(0),\g'(0))>0,\quad\Pi(\b'(\varepsilon),\g'(0))\Pi(\g'(0),\g'(0))\geq0.$$
$${\bf (H2)}\quad\quad\Pi(\b'(\varepsilon),\b'(\varepsilon))\Pi(\g'(0),\g'(0))>0,\quad\Pi(\b'(\varepsilon),\g'(0))\Pi(\g'(0),\g'(0))<0,$$
$$ \Pi(\zeta'(0),\g'(0)) \Pi(\g'(0),\g'(0))>0. $$
$${\bf (H3)}\quad\quad\Pi(\b'(\varepsilon),\b'(\varepsilon))\Pi(\g'(0),\g'(0))<0,\quad\Pi(\zeta'(0),\zeta'(0))\Pi(\g'(0),\g'(0))>0,$$
$$\Pi(\zeta'(0),\g'(0))\Pi(\g'(0),\g'(0))\geq0.$$
$${\bf (H4)}\quad\quad\Pi(\b'(\varepsilon),\b'(\varepsilon))
\Pi(\g'(0),\g'(0))<0,\quad\Pi(\zeta'(0),\zeta'(0))\Pi(\g'(0),\g'(0))<0,$$
$$\Pi(\zeta'(0),\b'(\varepsilon))\Pi(\zeta'(0),\zeta'(0))<0.$$

The notion of the noncharacteristic region is  a technical assumption, which may be not necessary for the solvability of (\ref{s}). In general, for $U\in T^2_\sym(S)$ given, there are many solutions to (\ref{s}). The aim of this assumption is to help us choose a regular solution for each $U.$ We shall solve problem (\ref{s}) locally in an asymptotic coordinate  and then paste the local solutions together, where the assumption of the noncharacteristic region is used to guarantee this produce to be successful. Case (I) is studied in \cite{Yao2017} where the density of smooth infinitesimal isometries in the $W^{2,2}$ infinitesimal isometries is obtained and the  matching property of infinitesimal
isometries  is proved.

If the middle surface is given by one single principal coordinate, that is, the assumption $(\ref{as})$ holds, then $S$ is in $(II)$ when for each $z\in[1,1+l],$ ${\bf r}(z,\cdot)$ is a closed
curve; otherwise, $S$  in $(I).$ However, there are   non-characteristic  regions which  can not be given by one single principal coordinate. For example, consider a surface, named as the monkey saddle, given by the graph of a function $h: $ $\R^2\rw\R,$
$$M=\{\,(x,h(x))\,|\,x=(x_1,x_2)\in\R^2\,\},$$ where $h=x_1^3-3x_1x_2^2.$ Then
$$ \kappa(x)=-\frac{36|x|^2}{(1+9|x|^4)^2}\qfq x=(x_1,x_2)\in\R^2.$$
For $b>1$ given, set
\be S=\{\,\a(t,s)\,|\,t\in[0,4),\,\,s\in[0,b]\,\},\quad\a(t,s)=\Big(\b(t,s),h(\b(t,s))\Big),\label{SS}\ee where
$$\b(t,s)=\left\{\begin{array}{l}(2b-s)\Big(1-t,\,\,\dfrac{t}{\sqrt{3}}\Big)\qfq t\in[0,2],\\
\Big(-(2b-s),\,\,\dfrac{2(2b-s)}{\sqrt{3}}(5-2t)\Big)\qfq t\in[2,3],\\
\Big((2b-s)(2t-7),\,\,-\dfrac{2(2b-s)}{\sqrt{3}}(4-t)\Big)\qfq t\in[3,4],
\end{array}\right.\qfq s\in[0,b].$$
A direct computation shows that the above $S$ belongs to (IV) such that the connection condition ({\bf H2}) holds at the connection points $t_1=0,$ $t_2=2,$ and $t_3=3,$ respectively.
We shall prove that there is no a single $\CC^1$ principal coordinate such that (\ref{as}) holds  as an appendix in the end of the paper.

Let $T^k$ denote the all $k$-order tensor fields on $S.$ Let $L^2(S,T^k)$ be the space of all $k$-order tensor fields on $S$ with the norm
$$(P_1,P_2)=\int_S\<P_1,P_2\>dg,$$ where
$$\<P_1,P_2\>=\sum_{i_1,\cdots,i_k=1}^2P_1(e_{i_1},\cdots,e_{i_k})P_1(e_{i_1},\cdots,e_{i_k})\qfq x\in S,$$ and $e_1,$ $e_2$ is orthonormal basis of $T_xS.$

We define $Q:$ $T_xM\rw T_xM$ by
\be Q\a=\<\a,e_2\>e_1-\<\a,e_1\>e_2\quad\mbox{for all}\quad\a\in T_xM,\label{qq}\ee where  $e_1,$ $e_2$ is an orthonormal basis of $T_xM$  with the positive orientation. Then the definition of $Q$ is independent of
the choice of a positively orientation orthonormal basis which is the rotation on $T_xM$ by $\pi/2$ along the clockwise direction, see \cite{Yao2017}.

To set up boundary data, we consider some boundary operators. Let $x\in\pl S$ be given. $\mu\in T_xS$ with $|\mu|=1$ is said to be the {\it noncharacteristic normal} outside $S$ if there is a curve
$\zeta:$ $(0,\varepsilon)\rw S$ such that
$$\zeta(0)=x,\quad \zeta'(0)=-\mu,\quad\Pi(\mu,X)=0\qfq X\in T_x(\pl S).$$ Let $\mu$ be the the noncharacteristic normal field along $\pl S.$
Recall that the shape operator $\nabla\n:$ $T_xM\rw T_xM$ is defined by $\nabla\n X=\nabla_X\n(x)$ for $X\in T_xM.$
 We define boundary operators $\T_i:$ $T_xM\rw T_xM$ by
\be \T_iX=\frac{1}{2}[X+(-1)^i\chi(\mu,X)\rho(X)Q\nabla\n X]\qfq X\in T_xM,\quad i=1,\,\,2,\label{xn4.14}\ee where
\be\chi(\mu,X)=\sign\det\Big(\mu,X,\n\Big),\quad \varrho(X)=\frac{1}{\sqrt{-\kappa}}\sign\Pi(X,X),\label{rho4.3}\ee and  $\sign$ is the sign function.

In (I) or (III), we shall consider the part boundary data
\be\<W,\T_1\a_s\>\circ\a(0,s)=q_1(s),\quad \<W,\T_1\a_s\>\circ\a(a,s)=q_2(s)\qfq s\in(0,b),\label{4.3}\ee
\be W\circ\a(t,0)=\phi\qfq t\in(0,a).\label{x1}\ee For convenience, we denote the relations (\ref{4.3}) and (\ref{x1}) by
\be W|_{I(III)(a,b)}=(q_1,\phi,q_2).\label{Ib}\ee

In (II) or (IV), the following boundary data are concerned
\be W\circ\a(t,0)=\phi\qfq t\in(0,a).\label{IIb}\ee

We have the following.
\begin{thm}\label{t1} $(i)$\,\,\, Let $S$ be given in $(I),$ or $(III)$ with the connection condition $({\bf H}1).$ Then there is a constant $C>0$ such that, for any $q_1,$ $q_2\in L^2(0,b),$ and $ \phi\in L^2((0,a),T)$ and  any $U\in L^2(S,T^2)$, there exists a unique solution $y$
to problem $(\ref{s})$ with the data $(\ref{Ib})$ satisfying
\be \|W\|_{L^2(S,T)}\leq C(\|U\|^2_{L^2(S,T^2)}+\|\phi\|^2_{L^2((0,a),T)}+\|q_1\|^2_{L^2(0,b)}+\|q_2\|^2_{L^2(0,b)}),\label{1.3}\ee where
$y=W+w\n.$

$(ii)$\,\,\,Let $S$ be given in $(II),$ or $(IV)$  with the connection condition $({\bf H}1).$ Then there is a constant $C>0$ such that, for any $ \phi\in L^2_a((0,a),T)$ and  any $U\in L^2(S,T^2)$, there exists a unique solution $y$
to problem $(\ref{s})$ with the data $(\ref{IIb})$
satisfying
\be \|W\|_{L^2(S,T)}\leq C(\|U\|_{L^2(S,T^2)}+\|\phi\|^2_{L^2_a((0,a),T)}),\label{1.4}\ee where
$y=W+w\n,$ and $L^2_a((0,a),T)$ is all $L^2$ vector fields defined in $\a(\cdot,0)$ with the period $a.$
\end{thm}

It follows immediately from Theorem \ref{t1} the corollary below

\begin{cor}\label{c1.1} Let $S$ be in $(I)$ or $(III)$  with the connection condition $({\bf H}1).$ Then there is a constant $C>0$ such that, for any $y=W+w\n\in H^1(S,\R^3)$ there exists an infinitesimal
$y^0\in H^1(S,\R^3)$ with the boundary data
$$W^0|_{I(III)}=\Big(\<W,\T_1\a_s\>\circ\a(0,\cdot),\,W\circ\a(\cdot,0),\,\<W,\T_1\a_s\>\circ\a(a,\cdot)\Big)$$
 satisfying
\be \|W-W^0\|_{L^2(S,T)}\leq C\|\Upsilon(y)\|_{L^2(S,T^2)},\label{1.8}\ee where $y^0=W^0+w^0\n.$

Let $S$ be given in $(II)$ or $(IV)$  with the connection condition $({\bf H}1).$ Then there is a constant $C>0$ such that, for any $y=W+w\n\in H^1(S,\R^3)$ there exists an infinitesimal
$y^0\in H^1(S,\R^3)$ with the boundary data
$$W^0\circ\a(\cdot,0)=W\circ\a(\cdot,0)$$ satisfying the estimate $(\ref{1.8}).$
\end{cor}

We have the following rigidity results in Theorems \ref{t1.2} and \ref{t1.3} on the strain tensor of the middle surface.

\begin{thm}\label{t1.2} $(i)$\,\,\, Let $S$ be given in $(I),$ or $(III).$  Then there is a constant $C>0$ such that
$$\|W\|^2_{L^2(S,T)}\leq C(\|\Upsilon(y)\|^2_{L^2(S,T^2)}+\|W\circ\a(\cdot,0)\|^2_{L^2((0,a),T)}+\|W\circ\a(0,\cdot)\|^2_{L^2((0,b),T)}+\|W\circ\a(a,\cdot)\|^2_{L^2((0,b),T)})$$
for  all $y=W+w\n\in H^1(S,\R^3).$

 $(ii)$\,\,\, Let $S$ be given in $(II),$ or $(IV).$  Then there is a constant $C>0$ such that
$$\|W\|^2_{L^2(S,T)}\leq C(\|\Upsilon(y)\|^2_{L^2(S,T^2)}+\|W\circ\a(\cdot,0)\|^2_{L^2((0,a),T)})$$
for  all $y=W+w\n\in H^1(S,\R^3).$
\end{thm}

\begin{thm}\label{t1.3} Let $S$ be a non-characteristic region from $(I)-(IV).$ Then
\be \|w\|^2_{L^2(S)}\leq C(\|Dw\|_{L^2(S)}\|\Upsilon(y)\|_{L^2(S)}+\|\Upsilon(y)\|^2_{L^2(S)})\label{1.12}\ee
for all $y=W+w\n\in H^1(S,\R^3)$
with $w|_{\pl S\times(-h/2,h/2)}=0$ and  $W|_{I(III)}=0$ or $W\circ\a(\cdot,0)=0,$ according to $S\in(I)\cup(III)$ or $S\in(II)\cup(IV),$ respectively.
\end{thm}

\begin{rem}
For the rigidity of the strain tensor, any of the connection conditions $({\bf H}1)-({\bf H4})$ is appropriate when the middle surface $S$ belongs to $(III),$ or $(IV).$
\end{rem}

We combine \cite[Theorem 3.1]{Ha3} with Theorem \ref{t1.2} by an argument as in \cite{Ha2} to obtain

\begin{thm}\label{t1.4} Let $S$ be a non-characteristic region from $(I)$-$(IV).$
Then  there are $C>0,$ $h_0>0,$ independent of $h>0,$ such that
\be\|\nabla y\|^2_{L^2(\Om)}\leq\frac{C}{h^{4/3}}\|\sym\nabla y\|^2_{L^2(\Om)}\label{1.13}\ee
for all $h\in(0,h_0)$ and $y=W+w\n\in H^1(\Om,\R^3)$ with $w|_{\pl S\times(-h/2,h/2)}=0$ and  $W|_{I(III)}=0,$ or $W|_{II(IV)}=0.$ Moreover, the exponential of the thickness $h$ in $(\ref{1.13})$ is optimal.
\end{thm}

\begin{rem} The results in Theorems $\ref{t1.2}$-$\ref{t1.4}$ are given in $\cite{Ha2}$ when the middle surface is given by one single principal coordinate.
\end{rem}

\setcounter{equation}{0}
\section{ A PDE system on $\R^2$}
\def\theequation{2.\arabic{equation}}
\hskip\parindent In an  asymptotic coordinate, equation (\ref{s}) can locally transfer to a PDE  system in (\ref{2.1}) below, see Proposition \ref{p4.1} later. Thus we study the solvability of (\ref{2.1}) in the present section which will be used in the next section to solve (\ref{s}) locally. Then we paste those local solutions together to
obtain a global one on $S.$

On $\R^2$  we consider the solvability of  problem
\be\left\{\begin{array}{l} f_{1x_1}(x)=a_{11}(x)f_1(x)+a_{12}(x)f_2(x)+p_1(x),\\
f_{2x_2}(x)=a_{21}(x)f_1(x)+a_{22}(x)f_2(x)+p_2(x),\end{array}\right.\qfq x=(x_1,x_2)\in\R^2,\label{2.1}\ee where  $(f_1,f_2)$ is the unknown,  $(p_1,p_2)$ is given, and $a_{ij}\in L^\infty.$

We shall work out some basic regions in which problem (\ref{2.1}) is uniquely solvable when $(p_1,p_2)$ and some data on part of boundary are given. Those regions are denoted by $E(\g),$ $R(z,a,b),$ $P_\pm(\b),$ $\Xi_\pm(\b,\g),$ and $\Phi(\b,\g,\hat\b),$ respectively. Their  definitions will be given in the following subsections.

A curve $\g(t)=(\g_1(t),\g_1(t)):$ $[0,t_0]\rw\R^2$ is said to be {\it noncharacteristic} if
$$\g'_1(t)\g'_2(t)\not=0\qfq t\in[a,b].$$

\subsection{Regions $E(\g)$ and $R(z,a,b)$}
\hskip\parindent
Let $\g(t)=(\g_1(t),\g_1(t)):$ $[0,t_0]\rw\R^2$ be a noncharacteristic curve such that
\be\g'_1(t)>0,\quad\g'_2(t)<0\qfq t\in(0,t_0).\label{non3.2}\ee   Set
\be E(\g) =\{\,(x_1,x_2)\in\R^2\,|\,\g_1\circ\g_2^{-1}(x_2)<x_1<\g_1(t_0),\,\g_2(t_0)<x_2<\g_2(0)\,\}.\label{T(Z,a)}\ee
Consider the boundary data
\be f\circ\gamma(t)=q(t)\qfq t\in(0,t_0).\label{2.33}\ee

For $\eta\in(0,t_0)$ fixed, consider the curve
$$\zeta(s)=\g(\eta)+s(1,1)\qfq s\in(0,s_0)$$ in $E(\g),$ where $s_0>0$ is such that $\zeta(s_0)\in\pl E(\g).$

Next, we consider a rectangle. For $z=(z_1,z_2)\in\R^2,$ $a>0,$ and $b>0$ given,  let
\be R(z,a,b)=(z_1,z_1+a)\times(z_2,z_2+b).\label{R}\ee
Consider the boundary data
\be f_1(z_1,x_2)=q_1(x_2),\quad f_2(x_1, z_2)=q_2(x_1)\label{xnn3.9}\ee for $x_1\in[z_1,z_1+a]$ and $x_2\in[z_2,z_2+b],$ respectively.

\begin{pro}\label{p2.1} For any $q=(q_1,q_2)\in L^2((0,t_0),\R^2)$ and $p=(p_1,p_2)\in L^2(E(\gamma),\R^2)$ given, there exists a unique solution $f=(f_1,f_2)\in L^2(E(\gamma),\R^2)$ to problem $(\ref{2.1})$
with the data $(\ref{2.33})$ satisfying
\be \|f\|^2_{L^2(E(\gamma),\R^2)}\leq C(\|q\|^2_{L^2((0,t_0),\R^2)}+\|p\|^2_{L^2(E(\gamma),\R^2)}),\label{2.5}\ee
\be\|f_1(\g_1(t_0),\cdot)\|^2_{L^2(\g_2(t_0),\g_2(0))}\leq C(\|q\|^2_{L^2((0,t_0),\R^2)}+\|p\|^2_{L^2(E(\gamma),\R^2)}),\label{2.555}\ee
\be\|f_2(\cdot,\g_2(0))\|^2_{L^2(\g_1(0),\g_1(t_0))}\leq C(\|q\|^2_{L^2((0,t_0),\R^2)}+\|p\|^2_{L^2(E(\gamma),\R^2)}),\label{2.55}\ee
\be\|f\circ\zeta\|^2_{L^2((0,s_0(\eta)),\R^2)}\leq C(\|q\|^2_{L^2((0,t_0),\R^2)}+\|p\|^2_{L^2(E(\gamma),\R^2)}),\label{2.552}\ee
\be \int_0^{t_0}[q_1\circ\g(t)|^2(t_0-t)+|q_2\circ\g(t)|^2t]dt\leq C(\|f\|^2_{L^2(E(\g),\R^2)}+\|p\|^2_{L^2(E(\g),\R^2)}).  \label{e211}    \ee
\end{pro}

\begin{pro}\label{p2x.2} For any $q_1\in L^2(z_2,z_2+b),$  $q_2\in L^2(z_1,z_1+a),$ and $p=(p_1,p_2)\in L^2(R(z,a,b),\R^2),$ problem $(\ref{2.1})$ admits a unique solution
$f=(f_1,f_2)\in L^2(R(z,a,b),\R^2)$ with the data $(\ref{R})$ satisfying
\be\|f\|^2_{L^2(R(z,a,b),\R^2)}\leq C(\|q_1\|^2_{L^2(z_2,z_2+b)}+\|q_2\|^2_{L^2(z_1,z_1+a)}+\|p\|^2_{L^2(R(z,a,b),\R^2)}).\label{px2.1}\ee
\end{pro}

The proofs of Propositions \ref{p2.1} and \ref{p2x.2} will be given after Lemma \ref{nl2.2}.

\begin{lem}\label{l2.1} Let $T>0$ be given. There is a $\varepsilon_T>0$ such that if $|\g(0)|\leq T$ and $\max\{\g_1(t_0)-\g_1(0),\g_2(0)-\g_2(t_0)\}<\varepsilon_T,$ then
the results in Proposition $\ref{p2.1}$ hold.
\end{lem}

{\bf Proof}\,\,\, The proof  is broken into several steps as follows.

{\bf Step 1.}\,\,\,
 Let $f=(f_1,f_2)$ solve problem (\ref{2.1}) and let $x=(x_1,x_2)\in E(\g)$ be given. We integrate the first equation in
(\ref{2.1}) with respect to the  variable $x_1$ over $(\g_1\circ\g_2^{-1}(x_2),x_1)$ where $x_2\in(\g_2(t_0),\g_2(0))$ is fixed to have
\be f_1(x_1,x_2)=q_1\circ\g\circ\g_2^{-1}(x_2)+\int_{\g_1\circ\g_2^{-1}(x_2)}^{x_1}(a_{11}f_1+a_{12}f_2+p_1)(\zeta_1,x_2)d\zeta_1.\label{2.57}\ee
Then integrating the second equation in (\ref{2.1}) over $(\g_2\circ\g_1^{-1}(x_1),x_2)$ with respect to the  variable $x_2$ yields
\be f_2(x_1,x_2)=q_2\circ\g\circ\g_1^{-1}(x_1)+\int_{\g_2\circ\g_1^{-1}(x_1)}^{x_2}(a_{12}f_1+a_{22}f_2+p_2)(x_1,\zeta_2)d\zeta_2.\label{2.56}\ee

{\bf Step 2.}\,\,\, We define an operator $\B:$ $L^2(E(\g),\R^2)\rw L^2(E(\g),\R^2)$ by
\beq &&\B f=\Big(q_1\circ\g\circ\g_2^{-1}(x_2),\,q_2\circ\g\circ\g_1^{-1}(x_1)\Big)+\Big(\int_{\g_1\circ\g_2^{-1}(x_2)}^{x_1}(a_{11}f_1+a_{12}f_2+p_1)(\zeta_1,x_2)d\zeta_1,\nonumber\\
&&\quad\quad\quad\quad\int_{\g_2\circ\g_1^{-1}(x_1)}^{x_2}(a_{12}f_1+a_{22}f_2+p_2)(x_1,\zeta_2)d\zeta_2\Big),\label{2.14}\eeq
for any $f=(f_1,f_2)\in L^2(E(\g),\R^2).$
 It is easy to check that
$f\in L^2(E(\g),\R^2)$ solves (\ref{2.1}) with the data (\ref{2.33}) if and only if $\B f=f.$

Next, we shall prove that there is a $0<\varepsilon_T\leq1$ such that when $|\g(0)|\leq T$ and $0<\max\{\g_1(t_0)-\g_1(0),\g_2(0)-\g_2(t_0)\}<\varepsilon_T,$ the map $\B:$ $L^2(E(\g),\R^2)\rw L^2(E(\g),\R^2)$
is contractible. Thus the existence and uniqueness of solutions follow from Banach's fixed point theorem.

In fact, for $f=(f_1,f_2),$ $\hat f=(\hat f_1,\hat f_2)\in L^2(E(\g),\R^2),$  it follows from (\ref{2.14}) that
\beq \B f-\B\hat f&&=\Big(\int_{\g_1\circ\g_2^{-1}(x_2)}^{x_1}[a_{11}(f_1-\hat f_1)+a_{12}(f_2-\hat f_2)](\zeta_1,x_2)d\zeta_1,\nonumber\\
&&\quad\int_{\g_2\circ\g_1^{-1}(x_1)}^{x_2}[a_{12}(f_1-\hat f_1)+a_{22}(f_2-\hat f_2)](x_1,\zeta_2)d\zeta_2\Big),\nonumber\eeq which yields
\beq|\B f-\B\hat f|^2&&\leq C_T\varepsilon[\int_{\g_1\circ\g_2^{-1}(x_2)}^{x_1}|f-\hat f|^2(\zeta_1,x_2)d\zeta_1+\int_{\g_2\circ\g_1^{-1}(x_1)}^{x_2}|f-\hat f|^2(x_1,\zeta_2)d\zeta_2],\nonumber\eeq
for $x=(x_1,x_2)\in E(\g),$ where $\varepsilon=\max\{\g_1(t_0)-\g_1(0),\g_2(0)-\g_2(t_0)\}.$ Thus we obtain
\be\|\B f-\B\hat f\|^2_{L^2}\leq C_T\varepsilon^2\|f-\hat f\|^2_{L^2},\label{n2.15}\ee i.e.,  the map $\B:$ $L^2(E(\g),\R^2)\rw L^2(E(\g),\R^2)$
is contractible if $\varepsilon$ is small.

{\bf Step 3.}\,\,\,
 Let  map  $\B:$ $L^2(E(\g),\R^2)\rw L^2(E(\g),\R^2)$  be defined in Step 2 and let $f\in L^2(E(\g),\R^2)$ be the solution to problem $(\ref{2.1})$ with the data
 (\ref{2.33}). It follows from (\ref{2.14}) and (\ref{n2.15}) that
 \beq\|f\|_{L^2(E(\g))}&&=\|\B f\|_{L^2(E(\g))}\leq \|\B(0)\|_{L^2(E(\g))}+\|\B f-\B(0)\|_{L^2(E(\g))}\nonumber\\
 &&\leq C_T(\|q\|_{L^2(0,t_0)}+\|p\|_{L^2(E(\g))})+C_T\varepsilon^2\|f\|_{L^2(E(\g))}.\nonumber\eeq Thus, the estimate (\ref{2.5}) follows if $\varepsilon$ is small. Moreover, it follows from
 (\ref{2.57}) that
 \beq|f_1(\g_1(t_0),x_2)|^2\leq C_T|q_1\circ\g\circ\g_2^{-1}(x_2)|^2+C_T\varepsilon\int_{\g_1\circ\g_2^{-1}(x_2)}^{\g_1(t_0)}(|f|^2+|p_1|^2)(\zeta_1,x_2)d\zeta_1, \nonumber\eeq
 for $x_2\in(\g_2(t_0),\g_2(0)),$ which yields the estimate (\ref{2.555}) by (\ref{2.5}). A similar argument gives (\ref{2.55}).

 Finally, we consider the estimate (\ref{2.552}). For $s\in[0,s_0]$ fixed we integrate the first equation in (\ref{2.1}) with respect to $x_1$ over $[\g_1\circ\g_2^{-1}(\g_2(\eta)+s),\g_1(\eta)+s]$ to have
 $$f_1\circ\zeta(s)=q_1(\g_2^{-1}(\g_2(\eta)+s))+\int_{\g_1\circ\g_2^{-1}(\g_2(\eta)+s)}^{\g_1(\eta)+s}(a_{11}f_1+a_{12}f_2+p_1)(\zeta_1,\g_2(\eta)+s)d\zeta_1,$$ which yields
 \beq|f_1\circ\zeta(s)|^2\leq2|q_1(\g_2^{-1}(\g_2(\eta)+s)|^2+C\int_{\g_1(0)}^{\g_1(t_1)}(|f_1|^2+|f_2|^2+|p_1|^2)(\zeta_1,\g_2(\eta)+s)d\zeta_1.\nonumber\eeq
 Thus we have
 $$\|f_1\circ\zeta\|^2_{L^2(0,s_0)}\leq C\|q_1\|^2_{L^2(0,t_0)}+C(\|f\|^2_{L^2(E(\g),\R^2)}+\|p\|^2_{L^2(E(\g),\R^2)}).$$
A similar argument gives
 $$\|f_2\circ\zeta\|^2_{L^2(0,s_0)}\leq C\|q_2\|^2_{L^2(0,t_0)}+C(\|f\|^2_{L^2(E(\g),\R^2)}+\|p\|^2_{L^2(E(\g),\R^2)}).$$ Thus the estimate (\ref{2.552}) follows from (\ref{2.5}) and the above inequalities.
 \hfill$\Box$

 By a similar argument as for Lemma \ref{l2.1} we have the following.

 \begin{lem}\label{nl2.2} Let $T>0$ be given. There is a $\varepsilon_T>0$ such that if $|z|\leq T$ and $\max\{a,b\}<\varepsilon_T,$ then
 Proposition $\ref{p2x.2}$ holds.
\end{lem}

{\bf Proof of Proposition \ref{p2.1}}\,\,\,\,We shall show that the assumptions£¬ $|\g(0)|\leq T$ and $\max\{\g_1(t_0)-\g_1(0),\g_2(t_0)-\g_2(0)\}<\varepsilon_T$ in Lemma \ref{l2.1} are unnecessary. Let $T>0$ be given such that
$$E(\g)\subset\{\,x\in\R^2\,|\,|x|\leq T\,\}.$$ Let $\varepsilon_T>0$ be given small such that Lemmas \ref{l2.1} and \ref{nl2.2} hold. We divide the curve $\g$ into
$m$ parts with the points $\tau_0=0,$ $\tau_0<\tau_1<\cdots<\tau_m=t_0$ such that
$$\quad |\g(\tau_{i+1})-\g(\tau_i)|=\frac{\varepsilon_T}{2},\quad 0\leq i\leq m-2,\quad |\g(t_0)-\g(\tau_{m-1})|\leq\frac{\varepsilon_T}{2}.$$
For simplicity, we assume that
$m=3.$ The other cases can be treated by a similar argument.

In the case of $m=3,$ we have
\beq \overline{E(\g)}&&=(\cup_{i=0}^2\overline{E}_i)\cup(\cup_{i=1}^3\overline{R}_i)\eeq where
$$E_i=\{\,x\in E(\g)\,|\,\g_1(\tau_i)< x_1<\g_1(\tau_{i+1}),\,\g_2(\tau_{i+1})< x_2<\g_2(\tau_i)\,\}\quad i=0,\,1,\,2,$$
$$R_1=[\g_1(\tau_1),\g_1(\tau_2)]\times[\g_2(\tau_1),\g_2(0)],\quad R_2=[\g_1(\tau_2),\g_1(t_0)]\times[\g_2(\tau_2),\g_2(\tau_1)],$$
$$R_3=[\g_1(\tau_2),\g_1(t_0)]\times[\g_2(\tau_1),\g_2(0)].$$

From Lemma \ref{l2.1}, problem (\ref{2.1}) admits a unique solution $f^i=(f^i_1,f^i_2)\in L^2(E_i,\R^2)$ for each $i=0,$ $1,$ and $2,$  respectively,
with the corresponding data and the corresponding estimates. We define $f\in L^2(\cup_{i=0}^2E_i,\R^2)$ by
$$f(x)=f^i(x)\qfq x\in \overline{E}_i\qfq i=0,\,1,\,2.$$ Then we extend the domain of $f$ from $\cup_{i=0}^3E_i$ to $E(\g)$ by the following way.
By Lemma \ref{nl2.2}, we define $f\in L^2(R_{i},\R^2)$
to be the solution $w^i=(w^i_1,w^i_2)\in  L^2(R_{i},\R^2)$ to problem (\ref{2.1}) with the data
$$w^i_1(\g_1(\tau_i),x_2)=f^i_1(\g_1(\tau_i),x_2)\qfq x_2\in[\g_2(\tau_{i}),\g_2(\tau_{i-1})],$$
$$w^i_2(x_1,\g_2(\tau_i))=f^i_2(x_1,\g_2(\tau_i))\qfq x_1\in[\g_1(\tau_i),\g_1(\tau_{i+1})], $$ for $i=1,$ and $2,$ respectively. Then we extend $f$ on $ L^2(R_3,\R^2)$ to be
the solution $w^3=(w^3_1,w^3_2)$ of (\ref{2.1}) with
the data
$$w^3_1(\g_1(\tau_2),x_2)=w^1_1(\g_1(\tau_2),x_2)\qfq x_2\in[\g_2(\tau_1),\g_2(0)],$$
$$w^3_2(x_1,\g_2(\tau_2))=w^2_2(x_1,\g_2(\tau_2))\qfq x_1\in[\g_1(\tau_2),\g_1(t_0)].$$

The estimates  in (\ref{2.5})-(\ref{2.55}) follow from the ones in Lemmas \ref{l2.1} and \ref{nl2.2}.

Moreover, the estimate (\ref{e211}) follows from the identities (\ref{2.57}) and (\ref{2.56}). \hfill$\Box$

A similar argument as above completes {\bf the proof of Proposition \ref{p2x.2}.}

\subsection{ Regions $P_\pm(\b)$}
\hskip\parindent
Let $\b=(\b_1,\b_2):$ $[0,t_1]\rw\R^2$ be a noncharcteristic curve such that
\be\b_1'(t)>0,\quad\b_2'(t)>0\qfq t\in[0,t_1].\label{xn3.4}\ee
Set
\be P_-(\b)=\{\,(x_1,x_2)\,|\,\b_1\circ\b_2^{-1}(x_2)<x_1<\b_1(t_1),\,\,\b_2(0)<x_2<\b_2(t_1)\,\},\label{x217}\ee
and consider the boundary data
\be \left\{\begin{array}{l}f_1\circ\b(t)=q_1(t)\qfq t\in(0,t_1);\\
 f_2(x_1,\b_2(0))=q_2(x_1)\qfq x_1\in(\b_1(0),\b_1(t_1)).\end{array}\right.\label{xnn3.11}\ee

Moreover,
set
\be P_+(\b)=\{\,(x_1,x_2)\,|\,\b_1(0)<x_1<\b_1\circ\b_2^{-1}(x_2),\,\,\b_2(0)<x_2<\b_2(t_1)\,\},\label{219}\ee
and consider the boundary data
\be \left\{\begin{array}{l}f_1(\b_1(0),x_2)=q_1(x_2)\qfq x_2\in(\b_2(0),\b_2(t_1));\\
 f_2\circ\b(t)=q_2(t)\qfq x_1\in(0,t_1).\end{array}\right.\label{xnn31.11}\ee

By a similar argument as for Proposition \ref{p2.1}, we have the following. The detailed proofs are omitted.

\begin{pro}\label{p2.2} For any $q_1\in L^2(0,t_1),$ $q_2\in L^2(\b_1(0),\b_1(t_1)),$ and $p=(p_1,p_2)\in L^2(E_-(\gamma),\R^2)$ given, there exists a unique solution $f=(f_1,f_2)\in L^2(P_-(\b),\R^2)$ to problem $(\ref{2.1})$
with the data $(\ref{xnn3.11}),$ such that
\be \|f\|^2_{L^2(P_-(\b),\R^2)}\leq C(\|q_1\|^2_{L^2(0,t_1)}+\|q_2\|^2_{L^2(\b_1(0),\b_1(t_1))}+\|p\|^2_{L^2(P_-(\b),\R^2)}).\label{2.10}\ee

\end{pro}

\begin{pro}\label{p2.22} For any $q_1\in L^2(\b_1(0),\b_1(t_1)),$ $q_2\in L^2(0,t_1),$ and $p=(p_1,p_2)\in L^2(E_+(\gamma),\R^2)$ given, there exists a unique solution $f=(f_1,f_2)\in L^2(P_+(\b),\R^2)$ to problem $(\ref{2.1})$
with the data $(\ref{xnn31.11}),$ such that
\be \|f\|^2_{L^2(P_+(\b),\R^2)}\leq C(\|q_1\|^2_{L^2(\b_1(\b_2(0),\b_2(t_1))}+\|q_2\|^2_{L^2(0,t_1)}+\|p\|^2_{L^2(P_+(\b),\R^2)}).\label{21.10}\ee
\end{pro}

\subsection{ Regions $\Xi_\pm(\b,\g)$}
\hskip\parindent
Let $\g=(\g_1,\g_2):$ $(0,t_0)\rw\R^2$ and $\b=(\b_1,\b_2):$ $(0,t_1)\rw\R^2$ be two noncharacterstic curves  with $\g(0)=\b(0)$ such that (\ref{non3.2}) and (\ref{xn3.4}) hold,
respectively.  We further assume that \be\b_1(t_1)\leq\g_1(t_0).\ee Set
\beq&&\Xi_-(\b,\g)=P_-(\b)\cup E(\g)\cup R\Big(z,a,b\Big),\label{set3.15}\eeq where $\P_-(\b),$  $E(\g),$ and $R(z,a,b)$ with $z=(\b_1(t_1),\b_2(0)),$ $a=\g_1(t_0)-\b_1(t_1),$ and $b=\b_2(t_1)-\b_2(0),$ are given in (\ref{x217}), (\ref{T(Z,a)}), and (\ref{R}), respectively.
Consider the boundary data
\be f_1\circ\b(t)=q_1(t)\qfq t\in(0,t_1),\label{x3.25}\ee
\be f\circ\gamma(t)=\hat q(t)\qfq t\in(0,t_0).\label{x3.26}\ee

Next, we assume that the  noncharacterstic curves  $\g=(\g_1,\g_2):$ $(0,t_0)\rw\R^2$ and $\b=(\b_1,\b_2):$ $(0,t_1)\rw\R^2$ with (\ref{non3.2}) and (\ref{xn3.4}), respectively, are given such that
$$\g(t_0)=\b(0),\quad \b_2(t_1)\leq\g_2(0).$$
Set
\be\Xi_+(\b,\g)=E(\g)\cup P_+(\b)\cup R(z,a,b),\ee where $E(\g),$ $P_+(\b),$ and $R(z,a,b)$  with $z=(\g_1(t_0),\b_2(t_1)),$ $a=\b_1(t_1)-\g_1(t_0),$ and $b=\g_2(0)-\b_2(t_1),$  are given in (\ref{T(Z,a)}), (\ref{219}), and (\ref{R}), respectively. Consider the boundary data
\be f_2\circ\b(t)=q_2(t)\qfq t\in(0,t_1),\label{x3.252}\ee
\be f\circ\gamma(t)=\hat q(t)\qfq t\in(0,t_0).\label{x3.262}\ee

Consider problem (\ref{2.1}) on the region $\Xi_-(\b,\g)$ with the boundary data (\ref{x3.25}) and (\ref{x3.26}). First, using Proposition \ref{p2.1}  we solve problem (\ref{2.1}) on the region $E(\g)$ with the data (\ref{x3.26}) to have a solution $f^1=(f^1_1,f^1_2)\in L^2(E(\g),\R^2).$ Then we solve problem (\ref{2.1}) on $P_-(\b)$ by Proposition \ref{p2.2}  using the data (\ref{x3.25}) and
$$f_1^2(x_1,\g_2(0))=f_1^1(x_1,\g_2(0))\qfq x_1\in(\g_1(0),\b_1(t_1))$$ to have a solution $f^2=(f^2_1,f^2_2)\in L^2(P_-(\b),\R^2).$ Next, by Proposition \ref{p2x.2}, we solve problem (\ref{2.1}) on the region $R(z,a,b)$ using
the data
$$ f^3_1(\b_1(t_1),x_2)=f^2_1(\b_1(t_1),x_2)\qfq x_2\in(\b_2(0),\b_2(t_1)),$$
$$f^3_2(\g_1(0),x_1)=f^1(\g_1(0),x_1)\qfq x_1\in(\b_1(t_1),\g_1(t_0)) $$ to have a solution  $f^3=(f^3_1,f^3_2)\in L^2(R(z,a,b),\R^2).$ Finally, we have a solution $f\in L^2(\Xi_-(\b,\g),\R^2)$ to problem (\ref{2.1}) with the data (\ref{x3.25}) and (\ref{x3.25}), given by
$$f=\left\{\begin{array}{l}f^1\qfq x\in E(\g),\\
f^2\qfq x\in P_-(\b),\\
f^3\qfq x\in R(z,a,b).\end{array}\right.$$
The above argument yields the following proposition where a detailed proof is omitted.
\begin{pro}\label{p2.35} For any $q_1\in L^2(0,t_1),$ $\hat q\in L^2((0,t_0),\R^2),$ and $p\in L^2(\Xi_-(\b,\gamma),\R^2)$ given, there is a unique solution $f=(f_1,f_2)\in L^2(\Xi_-(\b,\gamma),\R^2)$ to problem $(\ref{2.1})$
with the data $(\ref{x3.25})$ and $(\ref{x3.26})$ satisfying
\be \|f\|^2_{L^2(\Xi_-(\b,\gamma),\R^2)}\leq C(\|q_1\|^2_{L^2(0,t_1)}+\|\hat q\|^2_{L^2((0,t_0),\R^2)}+\|p\|^2_{L^2(\Xi_-(\b,\gamma),\R^2)}),\label{2.15}\ee
\be \|\hat q\circ\g(t)\|^2_{L^2((0,t_0-\varepsilon),\R^2)}\leq C_\varepsilon(\|f\|^2_{L^2(\Xi_-(\b,\gamma),\R^2)}+\|p\|^2_{L^2(\Xi_-(\b,\gamma),\R^2)}),\ee where $\varepsilon>0$ is small.
\end{pro}

By a similar argument as for Proposition \ref{p2.35}, we have the following. The detailed proof is omitted.
\begin{pro}\label{p2.36} For any $q_2\in L^2(0,t_1),$ $\hat q\in L^2((0,t_0),\R^2),$ and $p\in L^2(\Xi_+(\b,\gamma),\R^2)$ given, there is a unique solution $f=(f_1,f_2)\in L^2(\Xi_+(\b,\gamma),\R^2)$ to problem $(\ref{2.1})$
with the data $(\ref{x3.252})$ and $(\ref{x3.262})$ satisfying
\be \|f\|^2_{L^2(\Xi_+(\b,\gamma))}\leq C(\|q_2\|^2_{L^2(0,t_1)}+\|\hat q\|^2_{L^2((0,t_0),\R^2)}+\|p\|^2_{L^2(\Xi_+(\b,\gamma),\R^2)}),\label{2.152}\ee
\be \|\hat q\circ\g(t)\|^2_{L^2((0,t_0-\varepsilon),\R^2)}\leq C_\varepsilon(\|f\|^2_{L^2(\Xi_+(\b,\gamma),\R^2)}+\|p\|^2_{L^2(\Xi_+(\b,\gamma),\R^2)}),\label{e235}\ee where $\varepsilon>0$ is small.
\end{pro}

\subsection{ Region $\Phi(\b,\g,\hat\b)$}
\hskip\parindent
Let $\b$ and $\g$  be noncharacteristic curves  with $\b(0)=\g(0)$ and $\b_1(t_1)\leq\g_1(t_0)$ such that (\ref{non3.2}) and (\ref{xn3.4}) hold. Let $\hat{\b}=(\hat{\b}_1,\hat{\b}_2):$ $(0,\hat t_1)\rw\R^2$ be noncharacteristic  such that
$$\g(t_0)=\hat\b(0),\quad \hat\b_2(\hat t_1)\leq\g_2(0),\quad\hat\b_1'(t)>0,\quad\hat\b_2'(t)>0\qfq t\in[0,\hat t_1].$$ Set
\beq\Phi(\b,\g,\hat{\beta})&&=\Xi_-(\b,\g)\cup P_+(\hat\b)\cup R(z, a,b),\label{233}\eeq where $\Xi_-(\b,\g),$ $P_+(\hat\b),$ and $R(z,a,b)$ with $z=(\g_1(t_0),\hat\b_2(\hat t_1)),$ $a=\hat\b_1(\hat t_1)-\g_1(t_0),$ and $b=\b_2(t_1)-\hat\b_2(\hat t_1),$ are given in (\ref{set3.15}), (\ref{219}), and (\ref{R}), respectively.
Consider the boundary data
\be f_1\circ\b(t)=q_1(t),\quad t\in(0,t_1);\quad f_2\circ\hat\b(t)=q_2(t),\quad t\in(0,\hat t_1),\label{xx3.17}\ee
\be f\circ\gamma(t)=q(t)\qfq t\in(0,t_0).\label{xx3.18}\ee
By a similar argument as for Proposition \ref{p2.35}, we have the following. The detailed proof is omitted.
\begin{pro}\label{p2.4} Let $q_1,$ $q_2,$ and $q$ be given $L^2$ functions. Then problem $(\ref{2.1})$ admits a unique solution
$f=(f_1,f_2)\in L^2(\Phi(\b,\g,\hat\b),\R^2)$ with the data $(\ref{xx3.17})$ and $(\ref{xx3.18}).$
Moreover, the following estimates hold
$$\|f\|^2_{L^2(\Phi(\b,\g,\hat\b),\R^2)}\leq C(\|q\|^2_{L^2((0,t_0),\R^2)}+\|q_1\|^2_{L^2(0,t_1)}+\|q_2\|^2_{L^2(0,\hat t_1)}+\|p\|^2_{L^2(\Phi(\b,\g,\hat\b),\R^2)}).$$
\end{pro}

\setcounter{equation}{0}
\section{ Linear Strain Equations,  Proof of Theorem \ref{t1}}
\def\theequation{3.\arabic{equation}}
\hskip\parindent
We shall solve (\ref{s}) locally in asymptotic coordinate systems and then paste the local solutions together.   A chart $\psi(p)=(x_1,x_2)$ on $M$ is said to be an {\it asymptotic coordinate system} if
\be\Pi(\pl x_1,\pl x_1)=\Pi(\pl x_2,\pl x_2)=0.\label{3.4}\ee
 If $M$ is hyperbolic, an asymptotic coordinate system exists locally(\cite{Sp3}).

 Let $p\in M$ be given. Then there is an asymptotic coordinate
system $\psi:$ $\N\rw \R^2$ with $\psi(q)=(x_1,x_2)$ such that (\ref{3.4}) hold for $q\in\N,$ where $\N$ is a neighbourhood of $p.$  Let
$$G=\Big(g_{ij}(q)\Big),\quad g_{ij}=\<\pl x_i,\pl x_j\>.$$ Then
$$\Pi^2(\pl x_1,\pl x_2)=-\kappa\det G.$$

Our main observation is that in  an asymptotic coordinate system, equation (\ref{s}) takes the form (\ref{2.12}) below.

\begin{pro}\label{p4.1} Let $M$ be  a hyperbolic orientated surface and let  $\psi(p)=(x_1,x_2):$ $\N(\subset M)\rw\R^2$ be an asymptotic coordinate system  on $M$ with the positive orientation.   Then equation $(\ref{s})$ is equivalent to problem
\be\left\{\begin{array}{l}W_{1x_1}=\Ga_{11}^1W_1+\Ga_{11}^2W_2+U_{11},\\
W_{2x_2}=\Ga_{22}^1W_1+\Ga_{22}^2W_2+U_{22},\end{array}\right.\label{2.12}\ee where
$$W_i=\<W,\pl x_i\>,\quad U_{ij}=U(\pl x_i,\pl x_j),$$ and $\Ga_{ij}^k$ are the Christofell symbols for $1\leq i,\,\,j,\,\,k\leq2.$ Moreover, if $(W_1,W_2)$ solves problem $(\ref{2.12}),$ then $y=W+w\n$ is a solution to problem $(\ref{s})$ where
\be w\circ\psi^{-1}=\frac{1}{\omega}[U_{12}-\frac12(W_{1x_2}+W_{2x_1})+\Ga_{12}^1W_1+\Ga_{12}^2W_2],\quad \omega=\Pi(\pl x_1,\pl x_2).\label{2.13}\ee
\end{pro}

{\bf Proof}\,\,\,Problem (\ref{s}) is equivalent to
$$\Upsilon(y)(\pl x_i,\pl x_j)=U(\pl x_i,\pl x_j)\qfq 1\leq i,\,\,j\leq2.$$
Then the equations, $\Upsilon(y)(\pl x_i,\pl x_i)=U(\pl x_i,\pl x_i)$ for $i=1$ $2,$ yield problem (\ref{2.12}) since $\Pi(\pl x_i,\pl x_i)=0.$ In addition, (\ref{2.13}) follows from the equation $\Upsilon(y)(\pl x_1,\pl x_2)=U(\pl x_1,\pl x_2).$\hfill$\Box$

We also need the following lemmas \ref{l4.1}-\ref{l4.3}, whose proofs are given in \cite{Yao2017}.

\begin{lem}$(\cite{Yao2017})$\label{l4.1}
There is a $\si_0>0$ such that, for all $p\in \overline{S},$ there exist asymptotic coordinate systems $\psi:$ $B(p,\si_0)\rw\R^2$ with
$\psi(p)=(0,0),$
where $B(p,\si_0)$ is the geodesic plate in $M$ centered at $p$ with radius $\si_0.$
\end{lem}

\begin{lem}$(\cite{Yao2017})$\label{l4.2} Let $\gamma:$ $[0,a]\rw M$ be a regular curve without self intersection points. Then there is a $\si_0>0$ such that, for all $p\in\{\,\gamma(t)\,|\,t\in(0,a)\,\},$ $S(p,\si_0)$ has at most two
intersection points with $\{\,\gamma(t)\,|\,t\in[0,a]\,\},$ where $S(p,\si_0)$ is the geodesic circle centered at $p$ with radius $\si_0.$ If $\g(0)\not=\g(a),$  then  $S(p,\si_0)$
has at most
one intersection point with $\{\,\gamma(t)\,|\,t\in[0,a]\,\}$ for $p=\gamma(0),$ or $\gamma(a).$
\end{lem}

\begin{lem}$(\cite{Yao2017})$\label{l4.3} Let $p_0\in M$ and let $B(p_0,\si)$ be the geodesic ball centered at $p_0$ with radius $\si>0.$ Let $\gamma:$ $[-a,a]\rw B(p_0,\si)$ and $\b:$ $[-b,b]\rw B(p_0,\si)$ be two noncharacteristic curves of class $\CC^1,$ respectively, with
$$\gamma(0)=\b(0)=p_0,\quad \Pi(\dot\gamma(0),\dot\b(0))=0.$$
Let $\hat\psi:$ $B(p_0,\si)\rw\R^2$ be an  asymptotic coordinate. Then there exists an  asymptotic coordinate system  $\psi:$ $B(p_0,\si)\rw\R^2$
with $\psi(p_0)=(0,0)$ such that
\be\psi(\gamma(t))=(t,-t)\qfq t\in[-a,a],\label{ch4.27}\ee
\be\b_1'(s)>0,\quad \b_2'(s)>0\qfq s\in[-b,b],\label{4..16}\ee where
$\psi(\b(s))=(\b_1(s),\b_2(s)).$ Moreover, for $X=X_1\pl x_1+X_2\pl x_2$ with $\Pi(X,X)\not=0,$ we have
\be\varrho(X)Q\na\n X=\chi(\g'(0),\b'(0))\left\{\begin{array}{l} X_1\pl x_1-X_2\pl x_2,\quad X_1X_2>0,\\
-X_1\pl x_1+X_2\pl x_2,\quad X_1X_2<0,\end{array}\right.\label{x4.24}\ee where  $\varrho(X)$ and $Q$ are given in $(\ref{rho4.3})$ and $(\ref{qq}),$ respectively, and
$$\chi(\g'(0),\b'(0))=\sign\det\Big(\g'(0),\b'(0),\n(p_0)\Big).$$
\end{lem}

Let $S$ be given in (I). Let $0\leq t^-<t^+\leq a$ be fixed. Let $\zeta^\mp(s)=\a(\eta_\mp(s),s):$ $(0,\varepsilon)\rw S$ be noncharacteristic curves such that
$$\zeta^\mp(0)=\a(t^\mp,0),\quad{\zeta^\mp}'(0)\not=0,\quad\Pi(\a_t(t^\mp,0),{\zeta^\mp}'(0))=0.$$
Denote
\be S(\varepsilon)=\{\,\a(t,s)|\,t\in(\eta_-(s),\eta_+(s)),\,\,s\in(0,\varepsilon)\,\}.\label{1gg4.30}\ee
Consider the boundary data of $S(\varepsilon)$
\be\left\{\begin{array}{l}\<W,\T_1{\zeta^-}'\>\circ\zeta^-(s)=q_1(s),\quad \<W,\T_1{\zeta^+}'\>\circ\zeta^+(s)=q_2(s)\qfq s\in (0,\varepsilon),\\
W\circ\a(t,0)=\phi(t)\qfq t\in(t^-,t^+).\end{array}\right.\label{data3.8}\ee

\begin{lem}\label{l4.4}
There is a $\varepsilon>0$ small such that problem $(\ref{s})$ admits a unique
solution $y=W+w\n$ on $S(\varepsilon)$ with the  data $(\ref{data3.8})$  to satisfy
\be\|W\|^2_{L^2(S(\varepsilon),T)} \leq C(\|U\|^2_{L^2(S,T^2)}+\|q_1\|^2_{L^2(0,\varepsilon)}+\|\phi\|^2_{L^2((0,a),T)}+\|q_2\|^2_{L^2(0,\varepsilon)}),\label{xn4.24} \ee
\be\|\phi\|^2_{L^2((t_-,t_+),T)}\leq C(\|W\|^2_{L^2(S(\varepsilon),T)}+\|U\|^2_{L^2(S(\varepsilon),T^2)}).\label{310e}\ee
\end{lem}

{\bf Proof}\,\,\,Let $\si_0>0$ be given small such that the claims in Lemmas \ref{l4.1} and \ref{l4.2} hold, where $\gamma(t)=\a(t^-+t,0)$ for $t\in(0,t^+-t^-)$ in Lemma \ref{l4.2}.
We divide the curve $\g$ into $k$ parts with the points $\lam_i=\g(\tau_i)$  such that
$$\lam_0=\a(t^-,0),\quad \lam_k=\a(t^+,0),\quad d(\lam_i,\lam_{i+1})=\frac{\si_0}{3},\quad 0\leq i\leq k-2,\quad d(\lam_{k-1},\lam_k)\leq\frac{\si_0}{3},$$
where $\tau_0=0,$ $\tau_1>0,$ $\tau_2>\tau_1,$ $\cdots,$  and $\tau_k=t^+-t^->\tau_{k-1}.$ For simplicity, we assume that $k=3.$ The other cases can be treated by a similar argument.

We shall construct a local solution in a neighborhood of the curve $\g$ by the following steps.

{\bf Step 1.}\,\,\,Let $ s_0>0$ be small such that
$$\zeta^-(s)\in B(\lam_0,\si_0)\qfq s\in[0,s_0].$$
From Lemma \ref{l4.3}, there is  asymptotic coordinate system $\psi_0(p)=x:$ $B(\lam_0,\si_0)\rw\R^2$  with $\psi_0(\lam_0)=(0,0)$ such that
\be\g_0(t)=\psi_0(\g(t))=(t,-t)\qfq t\in[0,\tau_2],\label{4.16}\ee
\be\b_0(s)=\psi_0(\zeta^-(s))=(\b_{01}(s),\b_{02}(s)),\quad\b_{01}'(s)>0,\quad \b_{02}'(s)>0,\label{xp2.35}\ee for all $s\in[0,s_0].$
We may assume that $s_0>0$ small has been taken such that
$$\b_{01}(s_0)\leq\tau_2,$$ since $\psi(\zeta^-(0))=(0,0).$
Let the region $\Xi_-(\b_0,\g_0)\subset\R^2$ be given in (\ref{set3.15}) where $\b=\b_0$ and $\g=\g_0.$ Then set
$$S_0=S\cap\psi_0^{-1}[\Xi_-(\b_0,\g_0)].$$

Next, since ${\zeta^-}'(0)=\b_{01}'(s)\pl x_1+\b_{02}'(s)\pl x_2,$ from (\ref{x4.24}) and (\ref{xp2.35}), we have
$$\varrho({\zeta^-}')Q\na\n{\zeta^-}'(s)=\chi(\a_t(t^-,0),{\zeta^-}'(0))[\b_{01}'(s)\pl x_1-\b_{02}'(s)\pl x_2].$$ Noting that $\mu(\zeta^-(0))=-\a_t(t^-,0)/|\a_t(t^-,0)|$ and
$$\chi(\mu(\zeta^-(s)),{\zeta^-}'(s))=\chi(\mu(\zeta^-(0)),{\zeta^-}'(0))=-\chi(\a_t(t^-,0),{\zeta^-}'(0)),$$ from (\ref{xn4.14}), we obtain
\be\T_1{\zeta^-}'(s)=\b_{01}'(s)\pl x_1,\quad\T_2{\zeta^-}'(s)=\b_{02}'\pl x_2\qfq s\in(0,s_0).        \label{2.37}   \ee

Set
$$W_i=\<W,\pl x_i\>,\quad U_{ij}=U(\pl x_i,\pl x_j)\qfq 1\leq i,\,\,j\leq2,$$
$$\phi(t)=\phi_1(t)\pl x_1+\phi_2(t)\pl x_2.$$
From (\ref{2.37}) the boundary data $\<W,\T_1{\zeta^-}'\>\circ\zeta^-(s)=q_1(s)$ and $W\circ\a(t,0)=\phi(t)$ are equivalent to
\be W_1\circ\b_0(s)=\frac{q_1(s)}{\b_{01}'(s)},\quad(W_1,W_2)\circ\g_0(t)=\Big(\phi_1(t)g_{11}+\phi_2(t)g_{12},\,\phi_1(t)g_{12}+\phi_2(t)g_{22}\Big). \label{2.38}\ee
From Proposition \ref{p4.1}, the solvability of problem (\ref{s})  on $S_0$ with the boundary data (\ref{data3.8})
 is equivalent to
that of problem (\ref{2.12}) over the region $\Xi_-(\b_0,\g_0)$ with the boundary data (\ref{2.38}).

By Proposition \ref{p2.35}, problem (\ref{2.12}) admits a unique solution $(W_1,W_2)\in L^2(\Xi_-(\b_0,\g_0),\R^2)$ with the corresponding boundary data (\ref{2.38}).
 Thus, we have obtained a solution, denoted by
$y_0=W^0+w^0\n,$ to problem (\ref{s}) on
$S_0$ with the boundary data (\ref{data3.8}),
 where
$$W^0=(g^{11}W_1+g^{12}W_2)\pl x_1+(g^{12}W_1+g^{22}W_2)\pl x_2,\quad \Big(g^{ij}\Big)=G^{-1},$$ and $w^0$ is given by the formula (\ref{2.13}). Moreover, the inequality (\ref{2.15}) yields
the estimate
\beq\|W^0\|_{L^2(S_0,T)}^2\leq C(\|U\|_{L^2(S,T^2)}^2+\|q_1\|^2_{L^2(0,b)}+\|\phi\|^2_{L^2((0,a),T)}+\|q_2\|^2_{L^2(0,b)}).\label{xn4.281}\eeq

We define a curve on $S_0$ by
\be\zeta_1(s)=\psi_0^{-1}(s+\tau_1,s-\tau_1)\qfq s\in[0,s_{\tau_1}],\label{b4.32}\ee where
$$s_{\tau_1}=\left\{\begin{array}{l}\tau_1\quad\mbox{if}\quad \tau_1\leq\tau_2/2,\\
\tau_2-\tau_1\quad\mbox{if}\quad \tau_1>\tau_2/2.\end{array}\right. $$
Then $\zeta_1(s)$ is noncharacteristic and
\be\Pi(\zeta_1'(0),\a_t(\tau_1,0))=\Pi(\pl x_1+\pl x_2,\pl x_1-\pl x_2)=0.\label{xx4.17}  \ee
From (\ref{2.552}) the following estimate holds
\be\|W^0\circ\zeta_1\|_{L^2((0,s_{\tau_1}),T)}^2\leq C(\|U\|_{L^2(S,T^2)}^2+\|\phi\|^2_{L^2((0,a),T)}).\label{n252}\ee

{\bf Step 2.}\,\,\, Let the curve $\zeta_1$ be given in (\ref{b4.32}). Let $s_1>0$ be small such that
$$\zeta_1(s)\in B(\lam_1,\si_0)\qfq s\in[0,s_1].$$
From the noncharacteristicness of $\zeta_1(s)$ and the relation (\ref{xx4.17}) and Lemma \ref{l4.3} again, there exists  an  asymptotic coordinate system  $\psi_1(p)=x:$ $B(\lam_1,\si_0)\rw\R^2$ with $\psi_1(\lam_1)=(0,0)$ and
$$\g_1(t)=\psi_1(\a(t+\tau_1,0))=(t,-t)\qfq t\in[0,\tau_3-\tau_1],$$
$$\b_1(s)=\psi_1(\zeta_1(s))=(\b_{11}(s),\b_{12}(s)),\quad\b_{11}'(s)>0,\quad\b_{12}'(s)>0\qfq s\in[0,s_1].$$
 Since $\b_1(0)=(0,0),$ we may assume that $s_1>0$ is given such that
$$\b_{11}(s_1)\leq \tau_3-\tau_1.$$
Let $\Xi_-(\b_1,\g_1)$ be the region given in (\ref{set3.15}) where $\b=\b_1$ and $\g=\g_1.$ Set
$$S_1=S\cap\psi_1^{-1}[\Xi_-(\b_1,\g_1)].$$
As in (\ref{2.37}), we consider $\zeta_1(s)$  to be a part boundary of $S_1$ to have
$$\T_1\zeta_1'(s)=\b_{11}'(s)\pl x_1\qfq s\in(0,s_1).$$

 By a similar argument in Step 1, we obtain a unique solution
$y_1=W^1+w^1\n$ to problem (\ref{s}) on $S_1$ with the data
$$\<W^1,\T_1\zeta_1'\>\circ\zeta_1(s)=\<W^0,\T_1\zeta_1'\>\circ\zeta_1(s)\qfq s\in[0,s_1],$$
$$W\circ\a(t,0)=\phi\circ\a(t,0)\qfq t\in[\tau_1,\tau_3] ,$$ such that the following estimate holds
$$\|W^1\|_{L^2(S_1,T)}^2\leq C(\|U\|_{L^2(S,T^2)}^2+\|\<W^0,\T_1\zeta_1'\>\circ\zeta_1\|^2_{L^2(0,s_{\tau_1})}+\|\phi\|^2_{L^2((0,a),T)}).$$ Thus we obtain, by (\ref{n252}),
\be\|W^1\|_{L^2(S_1,T)}^2\leq C(\|U\|_{L^2(S,T^2)}^2+\|q_1\|^2_{L^2(0,b)}+\|\phi\|^2_{L^2(0,a),T)}+\|q_2\|^2_{L^2(0,b)})\label{n4.311}\ee

As in Step 1, we define a curve on $S_1$ by
$$\zeta_2(s)=\psi_1^{-1}(s+\tau_2-\tau_1,s+\tau_1-\tau_2)\qfq s\in[0,s_{\tau_2}],$$ where
$$  s_{\tau_2}=\tau_2-\tau_1\quad\mbox{if}\quad \tau_2-\tau_1\leq\frac{\tau_3-\tau_1}2;\quad  s_{\tau_2}=\tau_3-\tau_2\quad\mbox{if}\quad \tau_2-\tau_1>\frac{\tau_3-\tau_1}2.$$
 Then $\zeta_2(s)$ is noncharacteristic and
\be\Pi(\zeta_2'(0),\a_t(\tau_2,0))=\Pi(\pl x_1+\pl x_2,\pl x_1-\pl x_2)=0,\label{x4.17}  \ee and  from (\ref{2.552}) the following estimate holds
\be\|W^1\circ\zeta_2\|_{L^2((0,s_{\tau_2}),T)}^2\leq C(\|U\|_{L^2(S,T^2)}^2+\|\phi\|^2_{L^2((0,a),T)}).\label{321}\ee

{\bf Step 3.}\,\,\, Let $s_2>0$ be small such that
$$\zeta_2(s),\quad\a(a,s)\in B(\lam_2,\si_0)\qfq s\in[0,s_2].$$
Let $\psi_2(p)=x:$ $B(\lam_2,\si_0)\rw\R^2$ be an  asymptotic coordinate  with $\psi_2(\lam_2)=(0,0),$
\be\g_2(t)=\psi_2(\a(t+\tau_2,0))=(t,-t)\qfq t\in[0,t^+-t^--\tau_2],\label{4.27}\ee
$$\b_2(s)=\psi_2(\zeta_2(s))=(\b_{21}(s),\b_{22}(s)),\quad\b_{21}'(s)>0,\quad \b_{22}'(s)>0\qfq s\in[0,s_2].$$

Let  $\b_3(s)=\psi_2(\zeta^+(s))=(\b_{31}(s),\b_{32}(s)).$ Noting that $\b_3(0)=(t^+-t^--\tau_2,\tau_2-t^++t^-),$ we assume that $s_2$ has been taken small such that
$$\b_{32}(s_2)<0.$$
Next, we prove that
\be\b_{31}'(s)>0,\quad \b_{32}'(s)>0\qfq s\in[0,s_2],\label{4.28}\ee by contradiction.
Since $\b_3(s)$ is noncharacteristic, using (\ref{4.27}) and the assumption $\Pi(\a_t(t^+,0),\b_3'(0))=0,$ we have
$$\b_{31}'(0)=\b_{32}'(0);\quad\mbox{thus}\quad\b_{31}'(s)\b_{32}'(s)>0\qfq s\in[0, s_2].$$ Let
$$z(t,s)=\a_1(t,s)+\a_2(t,s),\quad\psi_2(\a(t+\tau_2,s))=(\a_1(t,s),\a_2(t,s)).$$
 Let (\ref{4.28}) be not true, that is,
$\b_{31}'(s)<0,$  $\b_{32}'(s)<0\qfq s\in[0, s_2].$ Thus
$$z(0,s)=\b_{21}(s)+\b_{22}(s)> \b_{21}(0)+\b_{22}(0)=0\qfq s\in(0,s_2],$$
$$z(t^+-t^-\tau_2,s)=\b_{31}(s)+\b_{32}(s)< \b_{31}(0)+\b_{32}(0)=0\qfq s\in(0,s_2].$$ Let $t(s)\in(0,t^+-t^-\tau_2)$ be such that
\be\a_1(t(s),s)+\a_2(t(s),s)=0\qfq s\in(0,s_2).\label{4.29}\ee Since $\a_{1t}(0,0)=1$ and $\a(t+\tau_2,s)$ are noncharacteristic for all $s\in[0,s_2],$ we have $\a_{1t}(t,s)>0$ and
$$0<\a_1(0,s)<\a_1(t(s),s)<\a_1(t^+-t^-\tau_2,s)=\b_{31}(s)<\b_{31}(0)=t^+-t^-\tau_2.$$ Thus, equality (\ref{4.29}) means that $\a(\a_1(t(s),s),0)=\a(t(s),s),$ which is a contradiction since
 $\a:$ $[0,a]\times[a,b]\rw M$ is an imbedding map.

 Let $\Phi(\b_2,\g_2,\b_3)$ be the region given in (\ref{233}) where $\b=\b_2,$ $\g=\g_2,$ and $\hat \b=\b_3.$ Set
 $$S_2=S\cap\psi_2^{-1}[\Phi(\b_2,\g_2,\b_3)].$$ Using (\ref{x4.24}) along the boundary of $S_2,$ we obtain
 $$\T_1\zeta_2'(s)=\b_{21}'(s)\pl x_1,\quad \T_1{\zeta^+}'(s)=\b_{32}'(s)\pl x_2\qfq s\in(0,s_2).$$

We again let
$$W_i=\<W,\pl x_i\>,\quad U_{ij}=U(\pl x_i,\pl x_j),$$ where $x=\psi_2.$ Thus problem (\ref{s}) on the region $S_2$ with the data
$$\<W,\T_1\zeta_2'\>\circ\zeta_2(s)=\<W^1,\T_1\zeta_2'\>\circ\zeta_2(s),\quad\<W,\T_1{\zeta^+}'\>\circ\zeta^+(s)=q_2(s)\qfq s\in(0,s_2),$$
$$W\circ\a(\tau_2+t,0)=\phi(\tau_2+t)\qfq t\in[0,t^+-t^--\tau_2],$$
is equivalent to that (\ref{2.12}) on $\Phi(\b_2,\g_2,\b_3)$ with the data
$$W_1\circ\b_2(s)=\frac{1}{\b_{21}'(s)}\<W^1,\T_1\zeta_2'\>\circ\zeta_2(s),\quad W_2\circ\b_3(s)=\frac{q_2(s)}{\b_{32}'(s)}\qfq s\in[0,s_2],$$ $$(W_1,W_2)(t,-t)=\Big(\phi_1(t)g_{11}+\phi_2(t)g_{12},\,\phi_1(t)g_{12}+\phi_2(t)g_{22}\Big)\qfq t\in[0,t^+-t^-\tau_2], $$
where $\phi=\phi_1\pl x_1+\phi_2\pl x_2$ and $y=W^1+w^1\n$ is the solution to problem (\ref{s}) on the region $S_1,$ given in Step 2.

We apply Proposition \ref{p2.4} with $f=(W_1,W_2)$ to  obtain a solution $y^2=W^2+w^2\n$ to problem (\ref{s}) on $S_2=S\cap\psi_2^{-1}(\Phi(\b_2,\g_2,\b_3)$ satisfies, by (\ref{321}),
\be\|W^2\|_{L^2(S_2,T)}^2\leq  C(\|U\|_{L^2(S,T^2)}^2+\|q_1\|^2_{L^2(0,b)}+\|\phi\|^2_{L^2((0,a),T)}+\|q_2\|^2_{L^2(0,b)}).\label{n4.312}\ee

{\bf Step 4.}\,\,\,We define
$$W=W^i,\quad w=w^i\qfq p\in S_i\qfq i=0,\,1,\,2.$$ Let $\varepsilon>0$ be small such that
$$S(\varepsilon)\subset S_0\cup S_1\cup S_2.$$
Then $y=W+w\n$ on $S(\varepsilon)$ will be a solution to (\ref{s}) with the corresponding boundary data if we show that
\be W^0(p)=W^1(p)\qfq p\in S_0\cap S_1;\quad W^1(p)=W^2(p)\qfq p\in S_1\cap S_2.\label{xx4.19}\ee
Since
$$\<W^1,\T_1\zeta_1'\>\circ\zeta_1(s)=\<W^0,\T_1\zeta_1'\>\circ\zeta_1(s)\qfq s\in[0,s_1],$$
$$W^1\circ\a(0,t)=\phi=W^0\circ\a(0,t)\qfq t\in[\tau_1,\tau_2],$$
from the uniqueness in Proposition \ref{p2.35}, we have
$$W^0\circ\psi_0^{-1}(x)=W^1\circ\psi_1^{-1}(x)\qfq x\in \Xi_-(\b_0,\g_0)\cap\Xi_-(\b_1,\g_1),$$ which yields the first identity in (\ref{xx4.19}).
A similar argument shows that the second identity in (\ref{xx4.19}) is true.

Finally, the estimate (\ref{xn4.24}) follows  from  (\ref{xn4.281}), (\ref{n4.311}), and (\ref{n4.312}).

{\bf Step 5}\,\,\,Let $y=W+w\n$ be a solution to problem (\ref{s}) on $S(\varepsilon)$ with the  data $(\ref{data3.8}).$ We now prove the estimate (\ref{310e}).

Let $t_0\in(t_-,t_+)$ be fixed.
Let $\psi:$ $B(\a(t_0,0),\si_0)\rw \R^2$ be an  asymptotic coordinate with $\psi(\a(t_0,0))=(0,0)$ such that
$$S\cap\psi^{-1}(E(\g))\subset S(\varepsilon),$$ where
$$\g(t)=\psi(\a(t_0+t,0))=(t,-t)\qfq t\in(-\si,\si)$$ and $0<\si<t_0-t_-$ is small. It follows from (\ref{e211}) that
$$\|\phi\|^2_{L^2((-\si/2,\si/2),T)}\leq C(\|W\|^2_{L^2(S(\varepsilon),T)}+\|U\|^2_{L^2(S(\varepsilon),T^2)}).$$

Next we consider similar estimates near the points $\a(t_-,0)$ and $\a(t_+,0).$ Let $\psi_0$ be the  asymptotic coordinate given in Step 1. Let $\si>0$ be small enough such that
$$\psi_0^{-1}(\Xi_-(\b_0,\g_0))\subset S(\varepsilon),$$ where $\b_0(s)$ and $\g_0(t)$ are given in (\ref{xp2.35}) and (\ref{4.16}) but their domains are $s\in(0,\si)$ and $t\in(0,\si),$ respectively.
From (\ref{e235}), we have
$$\|\phi\|^2_{L^2((t_-,t_-+\si/2),T)}\leq C(\|W\|^2_{L^2(S(\varepsilon),T)}+\|U\|^2_{L^2(S(\varepsilon),T^2)}).$$
By a similar argument, we obtain a similar estimate near the point $\a(t_+,0).$ Thus the estimate (\ref{310e}) follows from the finite covering theorem.
\hfill$\Box$

\begin{lem}\label{l35} Let $p=\a(t_i,0)$ be one of the connection points, given in $(III)$ or $(IV)$ with the connection condition $({\bf H}1).$
  Then there exist $\si>0$ and $C>0$  such that problem $(\ref{s})$ with the data $(\ref{x1})$ admits a unique solution $y=W+w\Pi$ on $B(p,\si)\cap S$ satisfying
\be\|W\|^2_{L^2(B(p,\si)\cap S,T)}\leq C(\|U\|^2_{L^2(S,T^2)}+\|\phi\|^2_{L^2((0,a),T)}).\label{e3.27}\ee
\end{lem}

{\bf Proof}\,\,\,Let
$$\b(t)=\a(t_i+t-\varepsilon,0),\quad \g(t)=\a(t_i+t,0)\qfq t\in[0,\varepsilon];\quad \zeta(t)=\a(t_i,t)\qfq t\in[-\varepsilon,\varepsilon],$$ where $\varepsilon>0$ is small.
Let $\psi:$ $B(p,\si)\rw\R^2$  an asymptotic coordinate with $\psi(p)=(0,0)$ such that
$$\psi(\g(t)))=(t,-t)\qfq t\in[0,\varepsilon]$$ for $\varepsilon>0$ small.
Let
$$\psi(\b(t))=(\b_1(t),\b_2(t))\qfq t\in[0,\varepsilon];\quad\psi(\zeta)=(\zeta_1(t),\zeta_2(t))\qfq t\in[-\varepsilon,\varepsilon].$$

Since $\Pi(\pl x_i,\pl x_i)=0$ for $i=1,$ $2,$ we have
$$\Pi(\b'(\varepsilon),\b'(\varepsilon))\Pi(\g'(0),\g'(0))=-4\b_1'(\varepsilon)\b_2'(\varepsilon)\Pi^2(\pl x_1,\pl x_2)(p),$$
$$\Pi(\b(\varepsilon),\g'(0))\Pi(\g'(0),\g'(0))=2[\b_1'(\varepsilon)-\b_2'(\varepsilon)]\Pi^2(\pl x_1,\pl x_2)(p).$$
Then the connection condition ({\bf H}1) in Section 1 implies that $\b_1'(t)\b_2'(t)<0$ for $t\in[0,\varepsilon]$ and $\b_1'(\varepsilon)-\b_2'(\varepsilon)\geq0.$ Thus
\be\b_1'(t)>0,\quad\b_2'(t)<0\qfq t\in[0,\varepsilon],\label{n3.31}\ee which means that the curve $(\b_1(t),\b_2(t))$ in Quadrant II since $(\b_1(\varepsilon),\b_2(\varepsilon))=\psi(p)=(0,0).$
We may assume that
\be S\cap\psi^{-1}(\mbox{\,Quadrant I\,})\not=\emptyset.\label{e332}\ee
Otherwise, we consider a new asymptotic coordinate $\hat\psi(q)=(-x_2,-x_1),$ where $(x_1,x_2)=\psi(q).$

Set
$$\Psi_1=E(\psi\circ\b)\cup R(0,\varepsilon,\b_2(0))\cup E(\psi\circ\g),$$ where
$$E(\psi\circ\g)=\{\,(x_1,x_2)\,|\,0<x_1<\varepsilon,\,-x_1<x_2<0\,\},\quad R(0,\varepsilon,\b_2(0))=(0,\varepsilon)\times(0,\b_2(0)),$$
$$E(\psi\circ\b)=\{\,(x_1,x_2)\,|\,\b_1\circ\b_2^{-1}(x_2)<x_1<0,\,0<x_2<\b_2(0)\,\}.$$
We apply Proposition \ref{p2.1} to $E(\psi\circ\b)$ and $E(\psi\circ\g)$  to obtain solutions $(W^0_1,W^0_2)\in L^2(E(\psi\circ\b),\R^2)$ and   $(W^1_1,W^1_2)\in L^2(E(\psi\circ\g),\R^2),$  respectively, to problem (\ref{2.12}) with the data
$$ (W^0_1,W^0_2)\circ\psi\circ\b(t)=\Big(\phi^0_1(t)g_{11}\circ\psi\circ\b(t)+\phi^0_2(t)g_{12}\circ\psi\circ\b(t),\,\,\phi^0_1(t)g_{12}\circ\psi\circ\b(t)+\phi^0_2(t)g_{22}\circ\psi\circ\b(t)\Big),$$
$$ (W^1_1,W^1_2)\circ\psi\circ\g(t)=\Big(\phi^1_1(t)g_{11}\circ\psi\circ\g(t)+\phi^1_2(t)g_{12}\circ\psi\circ\g(t),\,\,\phi^1_1(t)g_{12}\circ\psi\circ\g(t)+\phi^1_2(t)g_{22}\circ\psi\circ\g(t)\Big),$$
where $g_{ij}=\<\pl x_i,\pl x_j\>\circ\psi^{-1}(x),$ and
$$\phi(t)\circ\b(t)=\phi^0_1(t)\pl x_1+\phi^0_2(t)\pl x_2,\quad\phi(t)\circ\g(t)=\phi^1_1(t)\pl x_1+\phi^1_2(t)\pl x_2.$$ Then we have the unique solution $(W^2_1,W^2_2)\in L^2(R(0,\varepsilon_1,\b_2(0)),\R^2)$
to problem (\ref{2.12}) with the data
$$W^2_1(0,x_2)=W_1^0(0,x_2)\qfq x_2\in[0,\b_2(0)];\quad W^2_2(x_1,0)=W_2^1(x_1,0)\qfq x_1\in[0,\varepsilon_1].$$
Thus we have a solution $(W_1,W_2)\in L^2(\Psi_1,\R^2)$ to problem (\ref{2.12}) by the formula
$$(W_1,W_2)=(W_1^0,W^0_1)\qfq x\in E(\b);\quad(W_1,W_2)=(W_1^1,W^1_1)\qfq x\in E(\g);$$
$$ (W_1,W_2)=(W_1^2,W^2_1)\qfq x\in R(0,\varepsilon_1,\b_2(0)).$$ By Propositions  \ref{p2.1} and \ref{p2x.2}, $(W_1,W_2)\in L^2(\Psi,\R^2)$ is the solution to problem (\ref{2.12}) satisfying
\be\|(W_1,W_2)\|^2_{L^2(\Psi_1,\R^2)}\leq C(\|(\phi^0_1,\phi^0_2)\|^2_{L^2((0,\varepsilon_1),\R^2)}+\|(\phi^1_1,\phi^1_2)\|^2_{L^2((0,\varepsilon_1),\R^2)}+\|(U_{11},U_{22})\|^2_{L^2(\Psi_1,\R^2)}).\label{e3.29}\ee

Let
$$W=(g^{11}W_1+g^{12}W_2)\circ\psi(q)\pl x_1+(g^{12}W_1+g^{22}W_2)\circ\psi(q)\pl x_2\qfq q\in B(p,\si)\cup S,$$ where $(g^{ij})=(g_{ij})^{-1}$ and $\si>0$ is small enough. From (\ref{e3.29}) and Proposition \ref{p4.1},
$W$ is the solution to problem (\ref{s}) in $L^2(B(p,\si)\cup S,\R^2)$ with the data (\ref{x1}) satisfying  (\ref{e3.27}). \hfill$\Box$

Denote
\be S(0,s_0)=\{\,\a(t,s)|\,t\in(0,a),\,\,s\in(0,s_0)\,\}\qfq s_0\in[0,b].\label{gg4.30}\ee

\begin{lem}\label{l44}Let $S$ be given in $(I),$ or $(III)$ with the connection condition $({\bf H}1).$
Then there is a $0<\eta\leq b$ such that problem $(\ref{s})$ admits a unique
solution $y=W+w\n$ on $S(0,\eta)$ with the data  $(\ref{4.3})$ for $s\in(0,\eta)$ and $(\ref{x1})$ for $t\in(0,a)$ to satisfy
\be\|W\|^2_{L^2(S(0,\eta),T)} \leq C(\|U\|^2_{L^2(S,T^2)}+\|q_1\|^2_{L^2(0,b)}+\|\phi\|^2_{L^2((0,a),T)}+\|q_2\|^2_{L^2(0,b)}),\label{xn4.243} \ee
\be\|W\circ\a(\cdot,\eta)\|^2_{L^2((0,a),T)}\leq C(\|W\|^2_{L^2(S(0,\eta),T)}+\|U\|^2_{L^2(S(0,\eta),T^2)}).\label{3104e}\ee
\end{lem}

{\bf Proof}\,\,\,Let $S$ be given in (I). Set $\zeta^-(s)=\a(0,s)$ and $\zeta^+(s)=\a(a,s).$ Then the existence of the number $\eta>0$ and the estimate (\ref{xn4.243}) follow from  Lemma \ref{l4.4} and
(\ref{xn4.24}) immediately. Next, we let
$$\hat\a(t,0)=\a(t,\eta),\quad\hat\zeta^-(s)=\a(0,\eta-s),\quad\hat\zeta^+(s)=\a(a,\eta-s),\quad t_-=0,\quad t_+=a.$$ Then the estimate (\ref{3104e}) follows from (\ref{310e}).

We now suppose that $S$ is in (III) with the connection condition ({\bf H}1).

Different from the proof of Lemma \ref{l4.4}, we need to treat the  connection points $\a(t_i,s)$ for $1\leq i\leq m-1.$ For simplicity, we assume $m=3.$

Consider the connection points $p_i=\a(t_i,0)$ for $i=1,$ $2.$ By Lemma \ref{l35}, (a), there is $\si>0$ small such that problem (\ref{s}) admits a unique solution $y^i=W^i+w^i\n$ on $B(p_i,\si)\cap S$ with data (\ref{x1}) satisfying
the estimate (\ref{e3.27}).

 Let $\varepsilon>0$ be given small such that $\a(t_i\pm\varepsilon,0)\in B(p_i,\si).$ Let  $\zeta_i^\pm=\a(\eta_i^\pm(s),s):$ $(0,\varepsilon_1]\rw B(p_i,\si)\cap S$  be the noncharacteristic curves  such that
$$\zeta_i^\pm(0)=\a(t_i\pm\varepsilon,0),\quad\Pi(\a_t(t_i\pm\varepsilon,0),{\zeta_i^\pm}'(0))=0.$$
Set
$$S_0(\varepsilon_1)=\{\,\a(t,s)\,|\,t\in(0,\eta_1^-(s)),\,\,s\in(0,\varepsilon_1)\,\},$$
$$ S_1(\varepsilon_1)=\{\,\a(t,s)\,|\,t\in(\eta_1^+(s),\eta_2^-(s))),\,\,s\in(0,\varepsilon_1)\,\},$$
$$S_2(\varepsilon_1)=\{\,\a(t,s)\,|\,t\in(\eta_2^+(s)),a)\,\,s\in(0,\varepsilon_1)\,\}.$$

We apply Lemma \ref{l4.4} to the region $S_i(\varepsilon_1),$ respectively, for $i=0,$ $1,$ and $2.$ Then there is $0<\iota_0\leq\varepsilon_1$ such that problem $(\ref{s})$ admits a unique
solutions $y=W+w\n$ on $S_0(\iota_0)$ with the  data
$$W\circ\a(t,0)=\phi(t)\qfq t\in[0,t_1-\varepsilon],$$
$$\<W,\T_1\a_s\>\circ\a(0,s)=q_1(s),\quad\<W,\T_1{\zeta_1^-}'\>\circ\zeta_1^-(s)=\<W^1,\T_1{\zeta_1^-}'\>\circ\zeta_1^-(s)\qfq s\in[0,\iota_0].$$
We then obtain a unique
solution $y=W+w\n$  to problem $(\ref{s})$ on $S_1(\iota_1)$  for some $0<\iota_1\leq\varepsilon_1$ with the data
$$W\circ\a(t,0)=\phi(t)\qfq t\in[t_1+\varepsilon,t_2-\varepsilon],$$
$$\<W,\T_1{\zeta_1^+}'\>\circ\zeta_1^+(s)=\<W^1,\T_1{\zeta_1^+}'\>\circ\zeta_1^+(s),\quad\<W,\T_1{\zeta_2^-}'\>\circ\zeta_2^-(s)=\<W^2,\T_1{\zeta_2^-}'\>\circ\zeta_2^-(s),$$ for $s\in[0,\iota_1].$
Moreover, we  solve
 problem $(\ref{s})$ to have
solution $y=W+w\n$ on $S_2(\iota_2)$ for some $0<\iota_2\leq\varepsilon_1$ with the  data
$$W\circ\a(t,0)=\phi(t)\qfq t\in[t_2+\varepsilon,a],$$
$$\<W,\T_1{\zeta_2^+}'\>\circ\zeta_2^+(s)=\<W^2,\T_1{\zeta_2^+}'\>\circ\zeta_2^+(s),\quad\<W,\T_1\a_s\>\circ\a(a,s)=q_2(s)$$ for $s\in[0,\iota_2].$

Set $\eta>0$ small such that
$$S(0,\eta)\subset S_0(\iota_0)\cup B(p_1,\si)\cup S_1(\iota_1)\cup B(p_2,\si)\cup S_2(\iota_2).$$
By the uniqueness for problem (\ref{s}), we paste the above solutions together to obtain a unique solution to problem (\ref{s}) with the corresponding data such that (\ref{xn4.243}) and (\ref{3104e}) hold. \hfill$\Box$

\begin{lem}\label{l45}Let $S$ be given in  $(II),$ or  $(IV)$ with  the connection condition $({\bf H}1).$
Then there is a $0<\eta\leq b$ such that problem $(\ref{s})$ admits a unique
solution $y=W+w\n$ on $S(0,\eta)$ with the data  $(\ref{x1})$ for $t\in(0,a)$ to satisfy
\be \|W\|_{L^2(S(0,\eta),T)}\leq C(\|U\|_{L^2(S,T^2)}+\|\phi\|^2_{L^2_a((0,a),T)}),\label{e3.35}\ee
\be\|W\circ\a(\cdot,\eta)\|^2_{L^2((0,a),T)}\leq C(\|W\|^2_{L^2(S(0,\eta),T)}+\|U\|^2_{L^2(S(0,\eta),T^2)}),\label{336e}\ee where $S(0,\eta)$ is given in $(\ref{gg4.30}).$
\end{lem}

{\bf Proof}\,\,\,(1)\,\,\,Let $S$ be given in  $(II).$ Let $p_0=\a(0,0)=\a(a,0).$ For $\varepsilon>0$ small, set
$$\g(t)=\left\{\begin{array}{l}\a(t,0)\qfq t\in[0,\varepsilon),\\
\a(a+t,0)\qfq t\in(-\varepsilon,0).\end{array}\right.$$ Let
 $\psi:$ $B(p_0,\si_0)\rw\R^2$  be  an asymptotic coordinate such that
$$\psi(p_0)=(0,0),\quad\psi(\g(t))=(t,-t)\qfq t\in(-\varepsilon,\varepsilon).$$
By Proposition \ref{p2.1}, there is $0<\si\leq\si_0$ such that problem (\ref{s}) admits a unique solution $y=W^0+w^0\n$ on $B(p_0,\si)\cap S$ with the data
$W^0\circ\g(t)=\phi(t)$ for $t\in(-\varepsilon,\varepsilon)$ satisfying
$$\|W^0\|^2_{L^2(B(p_0,\si)\cap S,T)}\leq C(\|U\|^2_{L^2(S,T^2)}+\|\phi\|^2_{L^2((0,a),T)}).$$

Let $\zeta^\pm(s)=\a(\eta_\pm(s),s):$ $(0,\varepsilon_1)\rw S\cap B(p_0,\si)$ be noncharacteristic curves such that
$$\zeta^\pm(0)=\g(\pm\varepsilon),\quad\Pi({\zeta^\pm}'(0),\g'(\pm\varepsilon))=0.$$
We set
$$S_0=\{\,\a(t,s)\,|\,t\in(\eta_+(s),a+\eta_-(s)),\,\,s\in(0,\varepsilon_1)\}.$$ Then we solve problem $(\ref{s})$ on the region $S_0$ with the data
$$W\circ\a(t,0)=\phi(t)\qfq t\in[\varepsilon, a-\varepsilon],$$
$$\<W,\T_1{\zeta^+}'\>\circ\zeta^+(s)=\<W^0,\T_1{\zeta^+}'\>\circ\zeta^+(s),\quad\<W,\T_1{\zeta^-}'\>\circ\zeta^-(a,s)=\<W^0,\T_1{\zeta^-}'\>\circ\zeta^-(s)$$ for $s\in[0,\varepsilon_1],$ to obtain
the solution $y^1=W^1+w^1\n.$ Next, we paste the two solutions together to have the desired solution on $S(0,\eta)$ when $\eta$ is small.

(2)\,\,\, Let $S$ be given in (IV) with  the connection condition ({\bf H}1). Using Lemma \ref{l35}, (a), we treat the connection points, $\a(0,0),$ $\a(t_2,0),$ $\cdots,$ and $\a(t_{m-1},0),$ respectively, as in (1) to obtain the part solutions. Then we paste all the part solutions together to complete the proof. \hfill$\Box$\\

{\bf Proof of Theorem \ref{t1}}\,\,\,We suppose that $S$ is given in (I). The other cases can be treated by a similar argument. We omit the details.

Let $\aleph$ be the set of all $0<\eta\leq b$ such that the claims in Lemma \ref{l4.4} hold. We shall prove
$$b\in \aleph.$$

Let $\eta_0=\sup_{\eta\in\aleph}\eta.$  Then $0<\eta_0\leq b.$
Thus there is a unique solution $y=W+w\n$ on $S(0,\eta_0)$ to (\ref{s})  with the  data $W_{(I)(a,\eta_0)}=(q_1,\phi,q_2).$

As in the proof of Lemma \ref{l4.4}, we divide the curve $\a(\cdot,\eta_0)$ into $k$ parts with the points $\lam_i=\a(\tau_i,\eta_0)$  such that
$$\lam_0=\a(0,\eta_0),\quad \lam_k=\a(a,\eta_0),\quad d(\lam_i,\lam_{i+1})=\frac{\si_0}{3},\quad 0\leq i\leq k-2,\quad d(\lam_{k-1},\lam_k)\leq\frac{\si_0}{3},$$
where $\tau_0=0,$ $\tau_1>0,$ $\tau_2>\tau_1,$ $\cdots,$  and $\tau_k=a>\tau_{k-1},$ and $\si_0>0$ is given in Lemma \ref{l4.1}. For simplicity, we assume that $k=3.$

{\bf Step 1}\,\,\,
Let $\psi_0:$ $B(\lam_0,\si_0)\rw\R^2$ be an  asymptotic coordinate with $\psi_0(\lam_0)=(0,0)$ such that
$$\psi_0(\a(t,\eta_0))=(t,-t)\qfq t\in[0,\tau_2],$$
$$\psi_0(\a(0,\eta_0+s))=(\b_{01}(s),\b_{02}(s)),\quad\b_{01}'(s)>0,\quad\b_{02}'(s)>0\qfq s\in(-\varepsilon,\varepsilon),$$ for some $\varepsilon>0$ small. We take $\varepsilon_0>0$ such that
$$\g_0(t)=\psi_0(\a(t,\eta_0-\varepsilon_0)=(\g_{01}(t),\g_{02}(t))),\quad \g_{01}'(t)>0,\quad \g_{02}'(t)<0\qfq t\in[0,\tau_2],$$
$$\b_0(s)=\psi_0(\a(0,\eta_0-\varepsilon_0+s))=(\b_{01}(-\varepsilon_0+s),\b_{02}(-\varepsilon_0+s))\qfq s\in[0,2\varepsilon_0],\quad \b_{01}(\varepsilon_0)\leq\g_{01}(\tau_2).$$ Consider the region $\Xi_-(\b_0,\g_0),$ given in (\ref{set3.15}). Using Propositions \ref{p2.35} and \ref{p4.1}, we obtain a solution $y=W^0+w^0\n$ to problem (\ref{s}) on the region $S_0=S\cap\psi_0^{-1}(\Xi_-(\b_0,\g_0))$ with the data
$$\<W^0,\T_1\a_s\>\circ\a(0,\eta_0+s)=q_1(\eta_0+s)\qfq s\in(-\varepsilon_0,\varepsilon_0),$$
$$W^0\circ\a(t,\eta_0-\varepsilon_0)=W\circ\a(t,\eta_0-\varepsilon_0)\qfq t\in[0,\tau_2].$$ Moreover, it follows from (\ref{2.15}),  (\ref{3104e}), and (\ref{xn4.243}) that
\beq\|W^0\|^2_{L^2(S_0,T)}&&\leq C(\|U\|_{L^2(S,T^2)}^2+\|q_1\|^2_{L^2(0,b)}+\|W\circ\a(\cdot,\eta_0-\varepsilon_0)\|_{L^2((0,a),T)}^2)\nonumber\\
&&\leq C(\|U\|^2_{L^2(S,T^2)}+\|q_1\|^2_{L^2(0,b)}+\|\phi\|^2_{L^2((0,a),T)}+\|q_2\|^2_{L^2(0,b)}).\label{e336}\eeq
Set
$$\zeta_1(s)=\psi_0^{-1}((\tau_1,-\tau_1)+s(1,1))\qfq s\in(-\si_1,\si_1),$$ where $\si_1>0$ is small such that $\zeta_1(s)\in S$ for $s\in(-\si_1,\si_1).$ Clearly,
$\zeta_1$ is noncharacteristic to satisfy
$$\Pi({\zeta_1}'(0),\a_t(\tau_1,\eta_0))=0.$$ Moreover, it follows from (\ref{2.552}) and (\ref{e336}) that
$$\|W^0\circ\zeta_1\|^2_{L^2((-\si_1,\si_1),T)}\leq C(\|U\|^2_{L^2(S,T^2)}+\|q_1\|^2_{L^2(0,b)}+\|\phi\|^2_{L^2((0,a),T)}+\|q_2\|^2_{L^2(0,b)}).$$

{\bf Step 2}\,\,\,Let $\psi_1:$ $B(\lam_1,\si_0)\rw\R^2$ be an  asymptotic coordinate with $\psi_1(\lam_1)=(0,0)$ such that
$$\psi_1(\a(\tau_1+t,\eta_0))=(t,-t)\qfq t\in[0,a-\tau_1],$$
$$\psi_1(\zeta_1(s))=(\b_{11}(s),\b_{12}(s)),\quad\b_{11}'(s)>0,\quad\b_{12}'(s)>0\qfq s\in(-\si_1,\si_1).$$  
Let $0<\varsigma_{11}(\varepsilon_1)<\si_1$ and $0<\varsigma_{12}(\varepsilon_1)<a$ be given such that
$$\zeta_1(-\varsigma_{11}(\varepsilon_1))=\a(\varsigma_{12}(\varepsilon_1),\eta_0-\varepsilon_1)\qfq \varepsilon_1>0\quad\mbox{given small}.$$

We let $0<\varepsilon_1\leq\varepsilon_0$ be small such that
$$\g_1(t)=\psi_0(\a(\varsigma_{12}(\varepsilon_1)+t,\eta_0-\varepsilon_1)=(\g_{11}(t),\g_{12}(t))),\quad \g_{11}'(t)>0,\quad \g_{12}'(t)<0\qfq t\in[0,a-\varsigma_{12}(\varepsilon_1)],$$
$$\b_1(s)=(\b_{11}(-\varsigma_{11}(\varepsilon_1)+s),\b_{12}(-\varsigma_{11}(\varepsilon_1)+s))\qfq s\in[0,2\varsigma_{11}(\varepsilon_1)],\quad\b_{11}(\varepsilon_1)\leq\g_{11}(a-\varsigma_{12}(\varepsilon_1)).$$
We solve problem (\ref{s}) on the region $S_1=S\cup\psi_1^{-1}(\Xi_-(\b_1,\g_1))$ to have a solution $y=W^1+w^1\n$ with the data
$$\<W^1,\T_1{\zeta_1}'\>\circ\zeta_1(s)=\<W^0,\T_1{\zeta_1}'\>\circ\zeta_1(s) \qfq s\in[0,2\varsigma_{11}(\varepsilon_1))],$$
$$W^1\circ\a(\varsigma_{12}(\varepsilon_1)+t,\eta_0-\varepsilon_1)=W\circ\a(\varsigma_{12}(\varepsilon_1)+t,\eta_0-\varepsilon_1)\qfq t\in[0,a-\varsigma_{12}(\varepsilon_1)],$$ such that
\be\|W^1\|^2_{L^2(S_1,T)}\leq C(\|U\|^2_{L^2(S,T^2)}+\|q_1\|^2_{L^2(0,b)}+\|\phi\|^2_{L^2((0,a),T)}+\|q_2\|^2_{L^2(0,b)}).\ee

Set
$$\zeta_2(s)=\psi_1^{-1}((\tau_2,-\tau_2)+s(1,1))\qfq s\in(-\si_2,\si_2),$$ where $\si_2>0$ is small such that $\zeta_2(s)\in S$ for $s\in(-\si_2,\si_2).$
Then
$$\Pi({\zeta_2}'(0),\a_t(\tau_2,\eta_0))=0,$$
$$\|W^1\circ\zeta_2\|^2_{L^2((-\si_2,\si_2),T)}\leq C(\|U\|^2_{L^2(S,T^2)}+\|q_1\|^2_{L^2(0,b)}+\|\phi\|^2_{L^2((0,a),T)}+\|q_2\|^2_{L^2(0,b)}).$$

{\bf Step 3}\,\,\,As in Step 3 of the proof of Lemma \ref{l4.4}, there is an  asymptotic coordinate $\psi_2:$ $B(\lam_2,\si_0)\rw\R^2$ with $\psi_2(\lam_2)=(0,0)$ such that
\be\psi_2(\a(t+\tau_2,0))=(t,-t)\qfq t\in[0,a-\tau_2],\label{4.27}\ee
$$\psi_2(\zeta_2(s))=(\b_{21}(s),\b_{22}(s)),\quad\b_{21}'(s)>0,\quad \b_{22}'(s)>0\qfq s\in[-\si_2,\si_2],$$
$$\psi_2(\a(a,\eta_0+s))=(\b_{31}(s),\b_{32}(s)),\quad \b_{31}'(s)>0,\quad\b_{23}'(s)>0\qfq s\in[-\varepsilon_2,\varepsilon_2].$$ Let $0<\varsigma_{21}(\varepsilon_2)<\si_2$ and $0<\varsigma_{22}(\varepsilon_2)<a$ be given such that
$$\zeta_2(-\varsigma_{21}(\varepsilon_2))=\a(\varsigma_{22}(\varepsilon_2),\eta_0-\varepsilon_2)\qfq \varepsilon_2>0\quad\mbox{given small}.$$
Set
$$\g_2(t)=\psi_2(\a(\varsigma_{22}(\varepsilon_2)+t,\eta_0-\varepsilon_2))=(\g_{21}(t),\g_{22}(t))\qfq t\in[0,a-\varsigma_{22}(\varepsilon_2)],$$
$$ \b_2(s)=(\b_{21}(-\varsigma_{21}(\varepsilon_2)+s),\b_{22}(-\varsigma_{21}(\varepsilon_2)+s))\qfq s\in[0,2\varsigma_{21}(\varepsilon_2)],$$
$$\b_3(s)=(\b_{31}(\eta_0-\varepsilon_2+s),\b_{32}(\eta_0-\varepsilon_2+s))\qfq s\in[0,2\varepsilon_2].$$
Let $0<\varepsilon_2\leq\varepsilon_1$ be given small such that
$$\b_{21}(\varsigma_{21}(\varepsilon_2))\leq\g_{21}(a-\varsigma_{22}(\varepsilon_2)),\quad\b_{32}(\varepsilon_2)\leq \g_{22}(a-\varsigma_{22}(\varepsilon_2)),\quad\g_{21}'(t)>0,\quad\g_{22}'(t)<0.$$
Consider the region $\Phi(\b_2,\g_2,\b_2)$ given by (\ref{233}). We solve problem (\ref{s}) on the region $S_2=S\cup\psi_2^{-1}(\Phi(\b_2,\g_2,\b_3))$ to have a solution $y=W^2+w^2\n$ with the data
$$\<W^2,\T_1{\zeta_2}'\>\circ\zeta_2(s)=\<W^1,\T_1{\zeta_2}'\>\circ\zeta_2(s) \qfq s\in(0,2\varsigma_{21}(\varepsilon_2)),$$
$$\<W^2,\T_1\a_s\>\circ\a(a,\eta_0+s)=q_2(\eta_0+s)\qfq s\in[0,2\varepsilon_2],$$
$$W^1\circ\a(\varsigma_{22}(\varepsilon_2)+t,\eta_0-\varepsilon_2)=W\circ\a(\varsigma_{22}(\varepsilon_2)+t,\eta_0-\varepsilon_2)\qfq t\in[0,a-\varsigma_{22}(\varepsilon_2)],$$ such that
$$\|W^2\|^2_{L^2(S_2,T)}\leq C(\|U\|^2_{L^2(S,T^2)}+\|q_1\|^2_{L^2(0,b)}+\|\phi\|^2_{L^2((0,a),T)}+\|q_2\|^2_{L^2(0,b)}).$$

{\bf Step 4}\,\,\,Let $0<\varepsilon_3\leq \varepsilon_2$ be given small such that
$$S(0,\eta_0+\varepsilon_3)/S(0,\eta_0-\varepsilon_3)\subset S_0\cup S_1\cup S_2.$$ From Steps 1-3, we have extended the domain of the solution $y=W+w\n$ to the region $S(0,\eta_0+\varepsilon_3)$ such that
$$\|W\|^2_{L^2(S(0,\eta_0+\varepsilon_3)/S(0,\eta_0-\varepsilon_3),T)}\leq C(\|U\|^2_{L^2(S,T^2)}+\|q_1\|^2_{L^2(0,b)}+\|\phi\|^2_{L^2((0,a),T)}+\|q_2\|^2_{L^2(0,b)}),$$ which contradicts the definition of the number $\eta_0.$ \hfill$\Box$

\setcounter{equation}{0}
\section{ Rigidity,  Proofs of Theorems \ref{t1.2}-\ref{t1.4}}
\def\theequation{4.\arabic{equation}}
\hskip\parindent We need the following lemmas

\begin{lem} \label{l4.1} Let $p=\a(t_i,0)$ be one of the connection points, given in $(III)$ or $(IV)$ such that
one of  the connection conditions $({\bf H}2)$-$({\bf H}4)$ holds. Then there are $\si>0$ and $C>0$ such that
\be\|W\|^2_{L^2(B(p,\si)\cap S,T)}\leq C(\|\Upsilon(y)\|^2_{L^2(S,T^2)}+\|W\circ\a(\cdot,0)\|^2_{L^2((0,a),T)}),\label{e2.30}\ee for all $y=W+w\n\in H^1(S,\R^3).$
\end{lem}

{\bf Proof}\,\,\,Let
$$\b(t)=\a(t_i+t-\varepsilon,0),\quad \g(t)=\a(t_i+t,0)\qfq t\in[0,\varepsilon];\quad \zeta(t)=\a(t_i,t),$$ for $t\in[-\varepsilon,\varepsilon],$ where $\varepsilon>0$ is small.
Let $\psi:$ $B(p,\si)\rw\R^2$  an asymptotic coordinate with $\psi(p)=(0,0)$ such that
$$\psi(\g(t)))=(t,-t)\qfq t\in[0,\varepsilon]$$ for $\varepsilon>0$ small.
Let
$$\psi(\b(t))=(\b_1(t),\b_2(t))\qfq t\in[0,\varepsilon];\quad\psi(\zeta)=(\zeta_1(t),\zeta_2(t))\qfq t\in[-\varepsilon,\varepsilon].$$

Let the connection condition ({\bf H}2) in Section 1 hold. Then the inequalities
$$\Pi(\b'(\varepsilon),\b'(\varepsilon))\Pi(\g'(0),\g'(0))=-4\b_1'(\varepsilon)\b_2'(\varepsilon)\Pi^2(\pl x_1,\pl x_2)>0,$$
$$\Pi(\b'(\varepsilon),\g'(0))\Pi(\g'(0),\g'(0))=2[\b_1'(\varepsilon)-\b_2'(\varepsilon)]\Pi^2(\pl x_1,\pl x_2)<0,$$
$$ \Pi(\zeta'(0),\g'(0)) \Pi(\g'(0),\g'(0))=2[\zeta_1'(0)-\zeta_2'(0)]\Pi^2(\pl x_1,\pl x_2)>0$$
imply that
$$\b_1'(t)<0,\quad\b_2'(t)>0\qfq t\in[0,\varepsilon],\quad\cos\theta=\frac{1}{\sqrt{2[\zeta_1'^2(0)+\zeta_2'^2(0)]}}[\zeta_1'(0)-\zeta_2'(0)]>0,$$ where $\theta$ is the angular between $(1,-1)$ and $(\zeta_1'(0),\zeta_2'(0))$ in the plane $\R^2.$ Thus $\psi(B(p,\si)\cap S)$ is contained in Quadrant IV in the plane $\R^2$ if $\si>0$ is small enough.
Let
$$\Psi_2=E(\psi\circ\b(\varepsilon-\cdot))\cap E(\psi\circ\g),$$  where $E(\psi\circ\b(\varepsilon-\cdot))$ is given in (\ref{T(Z,a)}) with $\g(s)=\psi\circ\b(\varepsilon-s).$  Then
$$\Psi_2\subset\psi(B(p,\si)\cap S)\qfq\varepsilon>0\quad\mbox{small}.$$

Let $y=W+w\n\in H^1(S,\R^3)$ with $w=\<y,\n\>.$ Then $f=(\<W,\pl x_1\>,\<W,\pl x_2\>)$ solves problem (\ref{2.12}),  where $U_{ii}=\Upsilon(y)(\pl x_i,\pl x_i),$ on $\Psi_2$ with the corresponding data
on $\psi\circ\b$ and $\psi\circ\g.$ Thus it follows from Proposition \ref{2.1} that
$$\|W\|^2_{L^2(B(p,\si)\cap S,T)}\leq C\|f\|^2_{L^2(E(\psi\circ\b(\varepsilon-\cdot))\cap E(\psi\circ\g),\R^2)}\leq C(\|f\circ\psi\circ\b(\varepsilon-\cdot)\|^2_{L^2((0,\varepsilon),\R^2)}+\|f\circ\psi\circ\g\|^2_{L^2((0,\varepsilon),\R^2)}),$$ which imply that the estimate (\ref{e2.30}) is true.

Let one of the connection conditions ({\bf H}3) or ({\bf H}4) hold. A similar argument as above shows that estimate (\ref{e2.30}) holds. We omit the detail. \hfill$\Box$

\begin{lem}\label{l4.2} $(i)$\,\,\,Let $S$ be given in $(III)$ such that one of the connection conditions $({\bf H}2)-({\bf H}4)$ holds.
There are $\eta>0$ and $C>0$ such that for $y=W+w\n\in H^1(S,\R^3)$
\beq \|W\|^2_{L^2(S(0,\eta),T)}&&\leq C\Ga(\Upsilon,W,a,b),\label{e4.2}\eeq
 where $S(0,\eta)=\{\,\a(t,s)\,|\,(t,s)\in(0,a)\times(0,\eta)\,\}$ and 
$$\Ga(\Upsilon,W,a,b)=\|\Upsilon(y)\|^2_{L^2(S,T^2)}+\|W\circ\a(\cdot,0)\|^2_{L^2((0,a),T)}+\|W\circ\a(0,\cdot)\|^2_{L^2((0,b),T)}+\|W\circ\a(a,\cdot)\|^2_{L^2((0,b),T)}).$$ 
In addition, for any $0<\eta\leq b,$ the following estimate holds true
\be \|W\circ\a(\cdot,\eta)\|^2_{L^2((0,a),T)}\leq C(\|\Upsilon(y)\|^2_{L^2(S,T^2)}+\|W\|^2_{L^2(S(0,\eta),T)}).\label{e4.3}\ee

$(ii)$\,\,\,Let $S$ be given in $(IV)$ such that one of the connection conditions $({\bf H}2)-({\bf H}4)$ holds.
There are $\eta>0$ and $C>0$ such that  for $y=W+w\n\in H^1(S,\R^3)$
\beq \|W\|^2_{L^2(S(0,\eta),T)}&&\leq C(\|\Upsilon(y)\|^2_{L^2(S,T^2)}+\|W\circ\a(\cdot,0)\|^2_{L^2((0,a),T)}).\eeq Moreover, for any $0<\eta\leq b,$ the estimate $(\ref{e4.3})$ holds true.
\end{lem}

{\bf Proof}\,\,\,(i)\,\,\,We assume that $m=3.$ From Lemma \ref{l4.1}, there are $\si>0$ and $C>0$ such that
\be\|W\|^2_{L^2(B(p_i,\si)\cap S,T)}\leq C(\|\Upsilon(y)\|^2_{L^2(S,T^2)}+\|W\circ\a(\cdot,0)\|^2_{L^2((0,a),T)}),\label{ee4.2}\ee where $p_i=\a(t_i,0)$ for $i=1$ and $2.$

For $\varepsilon>0$ small, let
$$\g_0(t)=\a(t,0)\qfq t\in[0,t_1-\varepsilon];\quad \g_1(t)=\a(t,0)\qfq t\in[t_1+\varepsilon,t_2-\varepsilon];$$
$$\g_2(t)=\a(t,0)\qfq t\in[t_2+\varepsilon,a].$$ Next, we let $\zeta^\pm_i(s)=\a(\eta^i_\pm(s),s):$ $[0,\varepsilon]\rw S$ be given such that
$$\zeta_i^\pm(0)=\a(t_i\pm\varepsilon,0)\in B(p_i,\si),\quad\Pi({\zeta_i^\pm}'(0),\a_t(t_i\pm\varepsilon,0))=0\qfq i=1,\,\,2,$$ where $\varepsilon>0$ is given small enough. Using (\ref{xn4.24}) in Lemma \ref{l4.4}, we have
\be\|W\|^2_{L^2(S_i(\varepsilon_1),T)}\leq C\Ga(\Upsilon,W,a,b)\qfq i=0,\,\,1,\,\,2,\label{eee4.2}\ee where
$$S_0(\varepsilon)=\{\,\a(t,s)\,|\,t\in(0,\eta_-^1(s)),\,\,s\in(0,\varepsilon)\,\},\quad S_1(\varepsilon)=\{\,\a(t,s)\,|\,t\in(\eta_+^1(s),\eta_-^2(s)),\,\,s\in(0,\varepsilon)\,\},$$
$$S_2(\varepsilon)=\{\,\a(t,s)\,|\,t\in(\eta_+^2(s),a),\,\,s\in(0,\varepsilon)\,\}.$$

Let $\eta>0$ be small such that
$$S(0,\eta)\subset S_0(\varepsilon_1)\cup B(p_1,\si)\cup S_1(\varepsilon_1)\cup B(p_2,\si)\cup S_2(\varepsilon_2).$$ Then (\ref{e4.2}) follows from (\ref{ee4.2}) and (\ref{eee4.2}).

We now prove (\ref{e4.3}). Suppose that  the connection condition ({\bf H}2) holds. The other cases can be treated by a similar argument. Consider the connection points $p_i=\a(t_i,\eta)$ for $1\leq i\leq m.$ Let $\eta\in(0,b]$ be given.

Let
$$\b_i(t)=\a(t_i-\varepsilon+t,\eta),\quad \g_i(t)=\a(t_i+t,\eta)\qfq t\in[0,\varepsilon];\quad \hat\zeta_i(t)=\a(t_i,\eta-t),$$ for $t\in[-\varepsilon,\varepsilon],$ where $\varepsilon>0$ is small.
Let $\psi_i:$ $B(p_i,\si)\rw\R^2$  an asymptotic coordinate with $\psi_i(p_i)=(0,0)$ such that
$$\psi_i(\g_i(t)))=(t,-t)\qfq t\in[0,\varepsilon]$$ for $\varepsilon>0$ small.
Let
$$\psi_i(\b_i(t))=(\b_{i1}(t),\b_{i2}(t))\qfq t\in[0,\varepsilon];\quad\psi_i(\hat\zeta_i(t))=(\zeta_{i1}(t),\zeta_{i2}(t))\qfq t\in[-\varepsilon,\varepsilon].$$ 
Noting that $\hat\zeta_i(t)\in S(0,\eta)$ for $t\in(0,\varepsilon)$ and ${\hat\zeta}_i'(0)=-\zeta'(0),$ where $\zeta(s)=\a(t_i,\eta+t)$ is given in the definition of (III), as in the proof of Lemma \ref{l4.1}, the connection condition ({\bf H}2) implies that
$$\b_{i1}'(t)<0,\quad\b_{i2}'(t)>0\qfq t\in[0,\varepsilon],\quad\cos\theta=\frac{1}{\sqrt{2[\zeta_{i1}'^2(0)+\zeta_{i2}'^2(0)]}}[\zeta_{i1}'(0)-\zeta_{i2}'(0)]<0,$$ where $\theta$ is the angular between $(1,-1)$ and $(\zeta_{i1}'(0),\zeta_{i2}'(0))$ in the plane $\R^2.$ Thus $\psi_i(B(p_i,\si)\cap S(0,\eta))$ is contained in
$$ \mbox{Quadrant $I$}\cup\mbox{Quadrant $II$}\cup\mbox{Quadrant $III$}$$ in the plane $\R^2$ if $\si>0$ is small enough.
Let
$$\zeta_i^-(s)=\psi^{-1}(s(-\b_{i2}'(0),\b_{i1}'(0))),\quad \zeta_i^+(s)=\psi^{-1}(s(1,1))\qfq s\in[0,\si_0].$$ Then $\zeta_i^\pm(s)\in S(0,\eta)$ for $s\in[0,\si_0]$ when $\si_0>0$ is small, and
$$\Pi(\b'(\varepsilon),{\zeta_i^-}'(0))=0,\quad\Pi(\g'(0),{\zeta_i^+}'(0))=0.$$
Denote by $S_i$ the noncharacteristic region that consists of the curves $\zeta_{i-1}^+(s),$ $\a(t_i+t,\eta)$ for $t\in[0,t_{i+1}-t_i],$ and $\zeta_i^+(s).$ Clearly, 
$$S_i\subset S(0,\eta)\qfq 0\leq i\leq m-1,$$ when $\si_0>0$ is small enough. Applying the estimate (\ref{310e}) to the region $S_i,$ we obtain
$$\|W\circ\a(\cdot,\eta)\|_{L^2((t_i,t_{i+1}),T)}^2\leq C(\|W\|^2_{L^2(S_i,T)}+\|\Upsilon(y)\|^2_{L^2(S_i,T^2)})\qfq 0\leq i\leq m-1.$$
Thus the estimate (\ref{e4.3}) follows.  \hfill$\Box$\\

{\bf Proof of Theorem \ref{t1.2}}\,\,\,Using Lemma \ref{l4.2} and an argument as in the proof of Theorem \ref{t1}, we complete the proof. The details are omitted. \hfill$\Box$\\

{\bf Proof of Theorem \ref{t1.3}}\,\,\,  It follows from the  identity (\ref{2.52}) below that
\beq\int_Sw^2|\Pi|^2dg&&=\int_Sw\<\Pi,w\Pi\>dg=\int_S[w\<\Pi,\Upsilon(y)\>-w\<\Pi,DW\>]dg\nonumber\\
&&=-\int_{\pl S}w\Pi(W,\nu)d\pl S+\int_S[w\<\Pi,\Upsilon(y)\>+\Pi(W,Dw)+w\tr_g\ii(W)D\Pi]dg\nonumber\\
&&\leq C[\|w\|_{L^2(S)}(\|\Upsilon(y)\|_{L^2(S)}+\|W\|_{L^2(S)})+\|Dw\|_{L^2(S)}\|W\|_{L^2(S)}].\label{2.50}\eeq
Thus the estimate (\ref{1.12}) follows from (\ref{2.50}) and Theorem \ref{t1.2}.\hfill$\Box$

\begin{lem} For $(W,w)\in H^1(S,T)\times H^1(S),$ we have
\be\div_g[w\ii(W)\Pi]=\Pi(W,Dw)+w\tr_g\ii(W)D\Pi+w\<\Pi,DW\>\qfq p\in S.\label{2.52}\ee
\end{lem}

{\bf Proof}\,\,\,
Let $p\in S$ be given. Let $E_1,$ $E_2$ be a frame field normal at $p$ such that
$$\nabla_{E_i(p)}\n=\lam_iE_i(p),\quad D_{E_i(p)}E_j=0\qfq 1\leq i,\,\,j\leq2. $$
Thus we have at $p$
\beq\div_g[w\ii(W)\Pi]&&=E_1[w\Pi(W,E_1)]+E_2[w\Pi(W,E_2)]\nonumber\\
&&=E_1(w)\Pi(W,E_1)+E_2(w)\Pi(W,E_2)+wD\Pi(W,E_1,E_1)+wD\Pi(W,E_2,E_2)\nonumber\\
&&\quad+w\Pi(D_{E_1}W,E_1)+w\Pi(D_{E_2}W,E_2)\nonumber\\
&&=\Pi(W,Dw)+w\tr_g\ii(W)D\Pi+w\lam_1\<D_{E_1}W,E_1\>+w\lam_2\<D_{E_2}W,E_2\>\nonumber\\
&&=\Pi(W,Dw)+w\tr_g\ii(W)D\Pi+w\<\Pi,DW\>.\nonumber\eeq 

\setcounter{equation}{0}
\section{Optimal Exponential }
\def\theequation{5.\arabic{equation}}
\hskip\parindent We need an interpolation inequality from \cite{Ha3}. This result is also established in \cite{Yao2018} where  the existence of a local  principal coordinate is not assumed
but the Dirichlet boundary conditions are needed to hold on the thin faces of the shell.

\begin{thm}$(\cite{Ha3})$ Suppose that for each $p\in\overline{S}$ there exists locally a  principal coordinate at $p.$
Then  there are $C>0,$ $h_0>0,$ independent of $h>0,$ such that
\be\|\nabla y\|^2\leq C\Big(\frac{\|\<y,\n\>\|\|\sym\nabla y\|}h+\|y\|^2+\|\sym\nabla y\|^2\Big)\label{3.1}\ee
for all $h\in(0,h_0)$ and $y\in H^1(\Om,\R^3).$
\end{thm}

From \cite[Proposition 2,1]{Yao2018}, if $\kappa(p)<0,$ a local principal coordinate exists at $p.$ Thus, the estimates (\ref{3.1}) hold  when $S$ is a non-characteristic region.\\

By defining $\nabla\n\n=0,$ we introduce an 2-order tensor $p(y)$ on $\R^3_x$ by
\be p(y)(\tilde\a,\tilde\b)=\<\nabla_{\nabla\n\tilde\a}y,\tilde\b\>\qfq\tilde\a,\,\,\tilde\b\in\R^3.\label{3.2}\ee
Moreover, we need the following lemma from \cite{Yao2018}.

\begin{lem}$(\cite{Yao2018})$
 Let $y=W+w\n\in H^2(\Om,\R^3)$ be given. Then
\be|\nabla y+tp(y)|^2=|DW+w\Pi|^2+|Dw-\ii(W)\Pi|^2+|W_t|^2+w_t^2,\label{3.3}\ee
\be|\sym\nabla y+t\sym p(y)|^2=|\Upsilon(y)|^2+\frac12|X(y)|^2+w_t^2\qfq (p,t)\in S\times(-h/2,h/2),\label{3.4*}\ee where
$$\Upsilon(y)=\sym DW+w\Pi,\quad X(y)=Dw-\ii(W)\Pi+W_t.$$
\end{lem}

{\bf Proof of Theorem \ref{t1.3}}\,\,\,It follows from (\ref{3.2})-(\ref{3.4*}) that
$$(1-Ch)^2|\nabla y|^2\leq|\nabla y+tp(y)|^2\leq (1+Ch)^2|\nabla y|^2, $$
$$(1-Ch)^2|\sym\nabla y|^2\leq|\sym\nabla y+t\sym p(y)|^2\leq (1+Ch)^2|\sym\nabla y|^2. $$

From Theorem \ref{t1.3}, we have
\beq \|w\|^2_{L^2(S)}&&\leq C[(\|Dw-\ii(W)\Pi\|_{L^2(S)}+\|W\|_{L^2(S)})\|\Upsilon(y)\|_{L^2(S)}+\|\Upsilon(y)\|^2_{L^2(S)}]\nonumber\\
&&\leq C(\|\nabla y\|_{L^2(S)}\|\Upsilon(y)\|_{L^2(S)}+\|\Upsilon(y)\|^2_{L^2(S)}).\label{3.5}\eeq We integrate the above inequality in $t\in(-h/2,h/2)$ to have
$$ \|w\|\leq C(\sqrt{\|\nabla y\|\|\sym\nabla y\|}+\|\sym\nabla y\|).$$
Thus, by Holder's inequality, we obtain
\beq \frac{1}{h}\|w\|\|\sym\nabla y\|&&\leq C\frac{\|\nabla y\|^{1/2}\|\sym\nabla y\|^{3/2}}{h}+C\frac{\|\sym\nabla y\|^2}{h}\nonumber\\
&&= C(\varepsilon\|\nabla y\|^2)^{1/4}(\frac{\|\sym\nabla y\|^2}{\varepsilon^{1/3}h^{4/3}})^{3/4}+C\frac{\|\sym\nabla y\|^2}{h}\nonumber\\
&&\leq C(\frac{\varepsilon\|\nabla y\|^2}{4}+\frac{3}{4}\frac{\|\sym\nabla y\|^2}{\varepsilon^{1/3}h^{4/3}})+C\frac{\|\sym\nabla y\|^2}{h}\nonumber\\
&&\leq C\varepsilon\|\nabla y\|^2+C_\varepsilon\frac{\|\sym\nabla y\|^2}{h^{4/3}},\label{3.6}\eeq for $\varepsilon>0$ small.

In addition, from (\ref{3.5}) and Theorem \ref{t1.2}, we have
\be\|y\|^2\leq C\varepsilon\|\nabla y\|^2+C_\varepsilon\|\sym\nabla y\|^2,\label{3.7}\ee for $\varepsilon>0$ small.
Inserting (\ref{3.6}) and (\ref{3.7}) into (\ref{3.1}), we obtain (\ref{1.13}).

To complete the proof, we need to construct an Ansatz. From \cite[Proposition 2.1]{Yao2018}, there is a local principal coordinate on $S.$ In such a local principal coordinate, the Ansatz
has been given in \cite{Ha2}. \hfill$\Box$

\centerline{\bf Appendix:\,\,A Proof of that for $S$ in (\ref{SS}) there is}
\centerline{\bf no a single  principal coordinate such that (\ref{as}) holds true}
\def\theequation{A.\arabic{equation}}
\setcounter{equation}{0}
\vskip4mm

By contradiction.
We have
$$\pl x_1=(1,0,3(x_1^2-x_2^2)),\quad \pl x_2=(0,1,-6x_1x_2),$$
$$g=g_{11}dx_1^2+g_{12}(dx_1dx_2+dx_2dx_1)+g_{22}dx_2^2,$$
$$g_{11}=1+9(x_1^2-x_2^2)^2,\quad g_{12}=-18x_1x_2(x_1^2-x_2^2),\quad g_{22}=1+36x_1^2x_2^2,$$
\be\Pi(\a,\b)(p)=\si(x)(\a_1,\a_2)\left(\begin{array}{cc}-x_1&x_2\\
x_2&x_1\end{array}\right)\left(\begin{array}{c}\b_1\\
\b_2\end{array}\right),\quad \si(x)=\frac{6}{\sqrt{1+9|x|^4}},\label{A1}\ee
for $\a=\a_1\pl x_1+\a_2\pl x_2,$  $\b=\b_1\pl x_1+\b_2\pl x_2\in T_pS,$ $p=(x,h(x))\in S.$ The principal curvatures are the roots of the polynomial
\be(\lam g_{11}+\si x_1)(\lam g_{22}-\si x_1)=(\lam g_{12}-\si x_2)^2\qfq p=(x,h(x))\in S.\label{A2}\ee
Let the principal curvatures be $\lam_1>0>\lam_2$ for $ p=(x,h(x))\in S.$ It follows from (\ref{A2}) that
\be\lam_1(x_1,0)=\left\{\begin{array}{l}\si x_1\qfq x_1>0,\\
-\dfrac{\si x_1}{1+9x_1^4}\qfq x_1<0.\end{array}\right.\ee

Let $(z,\theta)$ be a principal coordinate of class $\CC^1$ such that (\ref{as}) holds.
We assume that
$$\lam_z=\lam_1\qfq p=(x,h(x))\in S.$$ Set
$$ E(x)=\frac{\pl z}{|\pl z|}=\zeta_1(x)\pl x_1+\zeta_2(x)\pl x_2\qfq (x,h(x))\in S.$$ Then $E(x)$ is globally defined on the whole $S$ such that
\be|E(x)|=1,\quad \nabla_{E(x)}\n=\lam_zE(x)\qfq (x,h(x))\in S.\label{A4}\ee
Next, we shall show that $\zeta_1$ and $\zeta_2$ can not be continuous simultaneously on the segments
$$\{\,(x_1,0)\,|\,x_1<-\frac{1}{\sqrt{3}}\,\}\quad\mbox{and}\quad \{\,(x_1,0)\,|\,x_1>0\,\}.$$ Thus a contradiction follows.

From (\ref{A1}) and (\ref{A4}), $(\zeta_1,\zeta_2)$ satisfies
$$ \lam_zg_{11}\zeta_1+\lam_zg_{12}\zeta_2=-\si x_1\zeta_1+\si x_2\zeta_2,\quad \lam_zg_{12}\zeta_1+\lam_zg_{22}\zeta_2=\si x_2\zeta_1+\si x_1\zeta_2\qfq (x,h(x))\in S, $$ from which we obtain
\be\zeta_1=\eta \zeta_2\qfq (x,h(x))\in S,\quad x_2\not=0,\label{A5}\ee where
$$\eta=\frac{\si|x|^2-\lam_z(g_{12}x_2+g_{22}x_1)}{\lam_z(g_{11}x_2+g_{12}x_1)}.$$
Using (\ref{A2}), we have
\be\eta=\frac{\lam_z(g_{11}g_{22}-g_{12}^2)+\si(g_{12}x_2-g_{11}x_1)}{\si(g_{11}x_2+g_{12}x_1)}\qfq (x,h(x))\in S,\quad x_2\not=0.\label{A7}\ee
It follows from (\ref{A4}) and (\ref{A5}) that
\be\zeta_2=\pm\frac{1}{\sqrt{\eta^2g_{11}+2\eta g_{12}+g_{22}}}\qfq (x,h(x))\in S,\quad x_2\not=0.\label{A8}\ee
By (\ref{A10}) below, in order for $\zeta_2$ to be continuous on the segment $\{\,(x_1,0)\,|\,x_1>0\,\}$ the sign in the right side hand of (\ref{A8}) must be the same for $x_2>0$ and $x_2<0.$ We assume that the sign in the right side hand of (\ref{A8}) is $+.$ Using (\ref{A5}) and (\ref{A9}) below, we have
$$\lim_{x_2\rw0^+}\zeta_1=-\frac1{\sqrt{1+9x_1^4}},\quad\lim_{x_2\rw0^-}\zeta_1=\frac1{\sqrt{1+9x_1^4}}\qfq x_1<-\frac1{\sqrt{3}},$$ respectively, which contradicts with the continuity of $\zeta_1$ on the segment $\{\,(x_1,0)\,|\,x_1<1/\sqrt{3}\,\}.$ \hfill$\Box$

A simple computation shows that the following lemma holds.

\begin{lem} Let $\eta$ be given by $(\ref{A7}).$ Then
\be \lim_{x_2\rw0} x_2\eta=\frac{-x_1(2+9x_1^4)}{1-9 x_1^4}\qfq x_1<-\frac1{\sqrt{3}},\label{A9}\ee
\be\lim_{x_2\rw0}\eta=0\qfq x_1>\frac{1}{\sqrt{3}}.\label{A10}\ee
\end{lem}

{\bf Compliance with Ethical Standards}

Conflict of Interest: The author declares that there is no conflict of interest.

Ethical approval: This article does not contain any studies with human participants or animals performed by the author.


\begin{thebibliography}{}



\bibitem{Ci} P. G. Ciarlet. Mathematical elasticity. Vol. III, volume 29 of Studies in Mathematics
and its Applications. North-Holland Publishing Co., Amsterdam, 2000. Theory of shells.


\bibitem{CiOlTr} D. Cioranescu, O. Oleinik, and G. Tronel. On Korn¡¯s inequalities for frame type structures
and junctions. C. R. Acad. Sci. Paris S¡äer. I Math., 309(9):591-596, 1989.

\bibitem{FrJaMu} G. Friesecke, R. James, S. Muller,  A theorem on geometric rigidity and the derivation of nonlinear
plate theory from three dimensional elasticity. Commun. Pure Appl. Math. 55, 1461-1506 (2002).

\bibitem{FrJaMu1}  G. Friesecke, R. James, S. Muller, A hierarchy of plate models derived from nonlinear elasticity by
gamma-convergence. Arch. Ration. Mech. Anal. 180(2), 183-236 (2006).

\bibitem{GeSa}G. Geymonat, E. Sanchez-Palencia, On the rigidity of certain surfaces with folds
and applications to shell theory. Arch. Ration. Mech. Anal. 129(1), 11-45 (1995).

\bibitem{GH} Y. Grabovsky and D. Harutyunyan,  Korn inequalities for shells with zero Gaussian curvature. Ann. Inst. H. Poincar¨¦ Anal. Non Lin¨¦aire  35  (2018),  no. 1, 267-282.

\bibitem{GH1}---,  Exact scaling exponents in Korn and Korn-type inequalities for cylindrical shells. SIAM J. Math. Anal.  46  (2014),  no. 5, 3277-3295.

\bibitem{Ha} D. Harutyunyan, On the Korn interpolation and second inequalities for shells with non-constant thickness,
arXiv:1709.04572 [math.AP].

\bibitem{Ha1}---, New asyptotically sharp Korn and Korn-like inequalities in thin domains.
Journal of Elasticity, 117(1), pp. 95-109, 2014.

\bibitem{Ha2}---, Gaussian curvature as an identifier of shell rigidity. Arch. Ration. Mech. Anal.  226  (2017),  no. 2, 743-766.

\bibitem{Ha3}---, On the Korn interpolation and second inequalities for shells with non-constant thickness,  arXiv:1709.04572 [math.AP].


\bibitem{HoLePa} P. Hornung, M. Lewicka, M. R. Pakzad,  Infinitesimal isometries on developable surfaces and asymptotic theories for thin developable shells. J. Elasticity 111 (2013), no. 1, 1-19.


\bibitem{LePa} M. Lewicka, M. R. Pakzad, The infinite hierarchy of elastic shell models: some recent
results and a conjecture. Infinite dimensional dynamical systems, 407-420, Fields Inst. Commun., 64, Springer, New York, 2013.



\bibitem{LeMoPa}  M. Lewicka, M. G. Mora, M. R. Pakzad,  Shell theories arising as low energy ¦£ -limit of 3d nonlinear elasticity. Ann. Sc. Norm. Super. Pisa Cl. Sci. (5) 9 (2010), no. 2, 253-295.

\bibitem{LeMoPa1}---, The matching property of infinitesimal isometries on elliptic surfaces and elasticity of thin shells. Arch. Ration. Mech. Anal. 200 (2011), no. 3, 1023-1050.

\bibitem{LeMu} M. Lewicka and S. Muller. On the optimal constants in korn¡¯s and geometric rigidity
estimates, in bounded and unbounded domains, under neumann boundary conditions.
 Indiana Univ. Math. J.  65  (2016),  no. 2, 377-397.

\bibitem{Lo} A. E. H. Love. A treatise on the mathematical theory of elasticity. Dover, 4th edition,
1927.

\bibitem{Na} S. A. Nazarov, Weighted anisotropic Korn¡¯s inequality for a junction of a plate and a
rod. Sbornik: Mathematics, 195(4):553-583, 2004.



\bibitem{Na1}---,  Korn inequalities for elastic junctions of massive bodies, thin plates, and
rods. Russian Mathematical Surveys, 63(1):35, 2008.

\bibitem{Ma} C. Mardare,  The generalized membrane problem for linearly elastic shells with hyperbolic or parabolic middle surface. J. Elasticity 51 (1998), no. 2, 145-165.

\bibitem{PaTo} R. Paroni and G. Tomassetti, Asymptotically exact Korns constant for thin cylindrical
domains. Comptes Rendus Mathematique, 350(15):749-752, 2012.

\bibitem{PaTo1} ---, On Korn¡¯s constant for thin cylindrical domains. Mathematics and Mechanics of Solids, 19(3):318-333, 2014.


\bibitem{Sa} E. Sanchez-Palencia,  Statique et dynamique des coques minces. II. Cas de flexion
pure inhibe¨¦. Approximation membranaire. C. R. Acad. Sci. Paris S¨¦r. I Math. 309(7),
531-537 (1989)

\bibitem{Sp3} M.  Spivak,  A comprehensive introduction to differential geometry. Vol. III. Second edition. Publish or Perish, Inc., Wilmington, Del., 1979. xii+466 pp. ISBN: 0-914098-83-7.

\bibitem{ToSm} Tovstik, P.E., Smirnov, A.L.: Asymptotic methods in the buckling theory of elastic
shells, volume 4 of Series on stability, vibration and control of systems.World Scientific,
Singapore 2001

\bibitem{Ho} H. Wu,  The Bochner Technique in Differential Geometry, Mathematical Re-
ports, Vol. 3, part 2, Harwood Academic Publishers, London-Paris, 1988.

\bibitem{Yao2012} P. F. Yao, Space of infinitesimal isometries and bending of shells, 2012,  arXiv:1310.5384.

\bibitem{Yao2011}---, Modeling and Control in Vibrational and Structural Dynamics. A differential geometric approach.
Chapman $\&$ Hall/CRC Applied Mathematics and Nonlinear Science Series. CRC Press, Boca Raton, FL, 2011.


\bibitem{Yao2017}---, Linear strain tensors on hyperbolic surfaces and asymptotic theories for
thin shells, arXiv:1708.07202 [math-ph].

\bibitem{Yao2018}---, Optimal exponentials of thickness in Korn's inequalities  for parabolic and elliptic shells,  arXiv:1807.11114 [math-ph].

\end{thebibliography}
 \end{document}